# PepC-Global: A Basin-Tuned Probabilistic Tropical Cyclone Model with Enhanced Out-of-Sample Skill and Climate-Sensitive Over-Land Decay

Cong Gao[1]*, Ning Lin[1]*

[1]Department of Civil and Environmental Engineering, Princeton University, USA

*Corresponding authors:

Cong Gao (cong.gao@princeton.edu)

Ning Lin (nlin@princeton.edu)

**Key Points:**

PepC-Global uniquely uses basin-wise tuning in a global tropical cyclone climatology framework.

PepC-Global genesis, track, and intensity modules outperform widely used linear models in out-of-sample evaluations.

PepC-Global intensity module uses environment-dependent over-land decay, offering greater climate change sensitivity than fixed decay rates.

## Abstract

We present PepC-Global, a global version of the Princeton environment-dependent probabilistic tropical cyclone (PepC) framework that uniquely implements basin-wise tuning within a single unified tropical cyclone climatology model. PepC-Global represents tropical cyclone climatology as a coupled stochastic process linking genesis, track, and intensity conditioned on large-scale environmental predictors. Each of the genesis, track, and intensity modules outperforms widely used linear models in out-of-sample tests. The intensity module also incorporates an environment-dependent over-land decay model, offering greater sensitivity to climate change signals than conventional fixed decay rate approaches. Systematic evaluation against observations demonstrates that PepC-Global closely reproduces genesis basin-wise frequency, seasonal cycles, interannual variability, and spatial distributions. The model also accurately captures basin-wise track patterns, along-coastline landfall frequency, and intensity statistics including lifetime maximum intensity and landfall intensity distributions. PepC-Global provides a versatile tool for probabilistic tropical cyclone hazard and risk assessment and a practical framework for investigating changes in tropical cyclone activity across future climate scenarios.

## Plain Language Summary

PepC-Global is a computer model designed to simulate tropical cyclones around the world in a probabilistic way, meaning it generates many plausible storms to represent the storm climatology. Unlike most existing global tropical cyclone climatology models, PepC-Global is tuned separately for each ocean basin within one unified framework, allowing it to better represent regional differences in where storms form, how they move, and how strong they become. Each of the genesis, track, and intensity modules outperforms widely used linear statistical models. Additionally, the intensity module includes an environment-dependent approach to represent how storms weaken over land, which better captures the influence of changing climate conditions compared to traditional methods that assume a fixed rate of weakening. Systematic comparison with historical observations shows that PepC-Global closely reproduces observed patterns of where and when storms form, how frequently they pass through different regions, where they make landfall, and how strong they become over their lifetimes and at landfall. PepC-Global can be used to estimate storm hazards and risks and provides a practical tool for studying how tropical cyclone activity may change in the future.

## 1. Introduction

Tropical cyclones (TCs) are among the most consequential natural hazards for society because they can combine destructive winds (Vickery et al., 2009), storm surge (Gori et al., 2022; N. Lin et al., 2012; Resio & Irish, 2015; Saffir, 1973; Xian et al., 2015), and heavy-rainfall-driven inland flooding (Gori et al., 2020; N. Lin & Shullman, 2017; Reed et al., 2015; Villarini et al., 2014; Yin et al., 2021) into a single event that strains infrastructure systems (Vanmarcke et al., 2013; Xian et al., 2018), threatens public health (Chu et al., 2025; Rice et al., 2025; Shultz et al., 2005; Z. Wang et al., 2025), and disrupts regional economies (Hsiang & Jina, 2014; Nordhaus, 2010; Pielke Jr et al., 2008; Weinkle et al., 2018). Hurricane Katrina (2005) underscores how these hazards can compound through cascading failures, as flooding linked to breakdowns in regional flood-protection systems drove widespread displacement and long, uneven recovery (Fussell, 2015; Kates et al., 2006; Seed et al., 2008). As TC exposure grows in many coastal regions under global warming (Hallegatte et al., 2013; Jing et al., 2024) and vulnerability varies across communities (Cutter et al., 2003), granular risk assessments are needed to inform equitable and effective adaptation strategies.

Decades of observational, theoretical, and modeling work show that the TC life cycle is primarily driven by large-scale environmental conditions, which dictate the storm's evolution alongside its internal dynamics. Genesis occurs preferentially over warm ocean waters of sufficient depth, with formation thresholds supported by both classic and modern analyses (Gao et al., 2022, 2025; Gray, 1968), and is most common between about 5° and 20° latitude, where the Coriolis effect can support vortex spin-up (Gray, 1968). In many basins, genesis is also linked to pre-existing synoptic-scale disturbances, including African easterly waves in the North Atlantic (NA; Hopsch et al., 2010) and monsoon-trough disturbances in the western North Pacific (WNP; Ritchie & Holland, 1999). After formation, storm motion is largely set by the environmental steering flow (Hall & Jewson, 2007) together with systematic drift associated with the meridional gradient of planetary vorticity (Fiorino & Elsberry, 1989). Intensity change is a function of maximum potential intensity (Emanuel, 1986, 2012; Holland, 1997; Ozawa & Shimokawa, 2015; Y. Wang et al., 2021), and it is strongly modulated by vertical wind shear (DeMaria & Kaplan, 1994), environmental moisture and ventilation (Tang & Emanuel, 2010), and upper-ocean thermal energy that can modulate storm-induced cooling (I.-I. Lin et al., 2008, 2009; Shay et al., 2000). Storm decay commonly follows landfall as friction increases and oceanic energy and moisture sources are cut off (Kaplan & DeMaria, 1995), and weakening can also occur when hostile atmospheric conditions promote dry-air intrusion (Fritz & Wang, 2013).

Despite these advances, substantial knowledge and modeling gaps remain in predicting TC genesis, track recurvature, and rapid intensification (RI), limiting our ability to anticipate risk-relevant extremes under global warming (Knutson et al., 2020; Walsh et al., 2016). Specifically, TC genesis frequency remains only weakly constrained by theory (Gao et al., 2025; Sobel et al., 2021). At the storm scale, there is still no first-principles theory that can deterministically predict which disturbances will develop into TCs (Emanuel, 2022; Sobel et al., 2021; Yang et al., 2021). The definition and diagnosis of the steering flow used to model storm motion vary across approaches and applications (J.-H. Chen et al., 2009; DeMaria et al., 2022; Galarneau & Davis, 2013; Hall & Jewson, 2007). Track recurvature is notoriously difficult to predict, and even modest differences in forecasts of when and where the turn begins can lead to substantially different storm trajectories and landfall outcomes (Barbero et al., 2024; Goerss, 2007). RI remains difficult to anticipate reliably even when large-scale conditions appear favorable (Kaplan et al., 2010; Rogers et al., 2013). These challenges are amplified under climate change, where nonstationarity, compounded by the coarse horizontal resolution and systematic biases of global climate models (e.g., Gao & Zhou, 2022), contributes to low consensus in simulating TC frequency, genesis patterns, track statistics, and intensity distributions (Camargo, 2013; Camargo et al., 2023; Knutson et al., 2020; Walsh et al., 2016).

To complement direct simulation of TCs, downscaling methods, including dynamical, statistical, and statistical-deterministic approaches, have been developed to overcome the coarse horizontal resolution and systematic biases of climate models. Because statistical and statistical-deterministic downscaling are far less computationally expensive than dynamical approaches, they can be used to generate very large samples of synthetic storms for hazard and risk assessment (Bloemendaal et al., 2020; Emanuel et al., 2006; Jing & Lin, 2020; Lee et al., 2018; Vickery et al., 2000). Vickery et al. (2000) introduced an empirical track model that initiates synthetic storms from historical data and propagates them using spatially varying linear regression models for track motion and intensity across NA. The Synthetic Tropical cyclOne geneRation Model (STORM) adopts a similar approach to Vickery et al. (2000) but extends it to global scale (Bloemendaal et al., 2020). Emanuel et al. (2006, 2008) advanced this framework by developing a statistical-deterministic model (hereafter KE08) in which randomly seeded incipient vortices are advected using a Beta and Advection Model (BAM) driven by large-scale steering flows, while intensity evolution is simulated by a dynamical model called the Coupled Hurricane Intensity Prediction System (CHIPS), enabling direct downscaling from reanalysis or climate model output. Note that random seeding does not produce an absolute rate of genesis and relies on observed genesis rates for calibration. Lee et al. (2018) developed an alternative statistical downscaling framework (now known as Columbia HAZard model or CHAZ) that replaces CHIPS with a linear regression intensity model, while also employing an environmental genesis index rather than random seeding. Jing and Lin (2020) introduced the Princeton environment-dependent probabilistic tropical cyclone (PepC) model, which departs from previous approaches by employing a random forest algorithm for track prediction and a hidden Markov model with covariates (Jing & Lin, 2019) for intensity evolution for NA. The PepC intensity module significantly improves over previous statistical models in capturing the observed distributions of intensity change, lifetime maximum intensity, and landfall intensity, particularly for extreme storms (Jing & Lin, 2019). Compared to KE08, PepC also better reproduces the spatial patterns of TC tracks, landfall frequency, and lifetime maximum intensity as driven by the High-Resolution Forecast-Oriented Low Ocean Resolution (HiFLOR) model (Jing et al., 2021). The success of PepC in reproducing observed TC genesis, track, and intensity statistics in NA motivates its extension to global scale to support TC hazard and risk assessment across all ocean basins.

A defining feature of TC climatology is strong basin specificity, as spatial variability is often dominated by interbasin and even sub-basin differences (T. Chen et al., 2008; Foltz et al., 2018; Guan et al., 2024), meaning models developed for one basin are not directly transferable to another. In NA, a large share of storms originate from African easterly waves (Hopsch et al., 2010), and many tracks recurve into the open ocean under the midlatitude westerlies (Hall & Jewson, 2007). In WNP, genesis frequently occurs within or near the monsoon trough (Molinari & Vollaro, 2013; Ritchie & Holland, 1999), and tracks can traverse long stretches of warm water that support intense storms and frequent landfalls in East Asia (Gao et al., 2022; I.-I. Lin & Chan, 2015). In the North Indian Ocean (NIO), TC activity exhibits pronounced pre-monsoon and post-monsoon peaks (Evan & Camargo, 2011; Li et al., 2013), with a sharp contrast between the Bay of Bengal (BoB) and the Arabian Sea (AS), where TC frequency in BoB can be roughly double that in AS. In the Southern Hemisphere, both the South Indian (SI) and South Pacific (SP) basins are active from November to April, with El Niño-Southern Oscillation (ENSO) strongly modulating genesis frequency and location (I.-I. Lin et al., 2020). Because a complete first-principles theory that predicts TC genesis, motion, and intensity across all regimes remains unavailable (Emanuel, 2018), these regional differences are not details to average away; they imply that a single global parameterization can be systematically misspecified for local hazard. Such basin-specific characteristics motivate basin-aware, and when needed sub-basin-aware, probabilistic models whose calibrated relationships and uncertainties adapt to regional dynamics.

Here, we upgrade PepC (Jing & Lin, 2020) from NA to a global, basin-tuned model, PepC-Global, that combines probabilistic environment dependence with basin-aware calibration to produce risk-relevant TC climatologies globally. PepC-Global differs from the original PepC in several ways that are specifically aimed at global applicability and basin-to-basin realism. First, the framework expands from a single-basin implementation to all major TC basins, while allowing basin-wise calibration so regionally distinct relationships are retained rather than forced into a single global fit. Second, the genesis module is upgraded from the widely used Poisson linear regression (Jing & Lin, 2020; Tippett et al., 2011) to a support vector machine (SVM) classifier, which is better suited to capturing nonlinear, threshold-like environmental controls on genesis (Gray, 1968; Tippett et al., 2011). Third, we revise the track module to remove reliance on analog selection and to include an explicit latitude dependence that captures systematic turning associated with beta drift, improving portability across basins while still representing observed track statistics. Fourth, the intensity module incorporates an environment-dependent over-land decay model, offering greater sensitivity to climate change signals than fixed decay rate approaches (Jing & Lin, 2019; Kaplan & DeMaria, 1995, 2003). Building on these component updates, PepC-Global simulates genesis, track, and intensity as a coupled stochastic process conditioned on large-scale predictors, and evaluates both accuracy and probabilistic reliability across basins and metrics.

This paper is organized as follows. Section 2 describes the observational and reanalysis datasets, the three-module PepC-Global framework (genesis, track, and intensity), and the experimental design used to evaluate each module individually and in fully coupled mode. Section 3 presents the PepC-Global genesis module and evaluates the simulated seasonal cycle, interannual variability, and the spatial distribution of genesis across basins. Section 4 introduces the PepC-Global track module and evaluates the simulated basin-wise track passage patterns and along-coastline landfall frequency. Section 5 describes the PepC-Global intensity module and evaluates the simulated lifetime maximum intensity and landfall intensity distributions. Section 6 evaluates the fully coupled PepC-Global system. Section 7 summarizes the main findings and discusses implications and potential applications of PepC-Global for TC hazard and risk analysis.

## 2. Data and Methods

### 2.1 Data

To develop PepC-Global, we obtain atmospheric fields and sea surface temperature (SST) from fifth-generation European Centre for Medium-Range Weather Forecasts (ECMWF) atmospheric reanalysis (ERA5), and oceanic subsurface fields from the US National Oceanic and Atmospheric Administration (NOAA) National Centers for Environmental Prediction (NCEP) Global Ocean Data Assimilation System (GODAS). ERA5 provides a globally complete, observation-constrained record of atmospheric and surface conditions produced through data assimilation within a numerical weather prediction system (Hersbach et al., 2020). It is distributed on regular latitude-longitude grids at 0.25° × 0.25° resolution, which corresponds to roughly 31 km at the equator, and includes both single-level and standard pressure-level fields. The core ERA5 product provides hourly estimates for a wide range of variables, and companion products provide monthly means and post-processed daily statistics derived from the hourly data. GODAS provides gridded global ocean analyses on a 0.333° latitude × 1° longitude grid (Behringer et al., 1998), and is commonly distributed as pentad (5-day) and monthly mean fields.

Large-scale predictors are derived from monthly ERA5 and GODAS fields and then regridded to a common 2.5° × 2.5° grid for input to PepC-Global. From monthly ERA5, we derive atmospheric predictors that describe the dynamical and thermodynamic environment for TC activity, including relative vorticity at 850 hPa, horizontal winds at 200, 250, and 850 hPa, and relative humidity at 600 hPa. Potential intensity is computed from hourly ERA5 thermodynamic fields (air temperature, specific humidity, SST, and mean sea level pressure) following established formulations (Bister & Emanuel, 2002), and is then aggregated to monthly values prior to use. For the over-land decay model, we additionally obtain volumetric soil water content and soil temperature from ERA5. From monthly GODAS seawater potential temperature, we estimate mixed layer depth and upper-ocean stratification, and combine these ocean metrics into an ocean feedback factor that captures the expected suppression of intensification by storm-induced cooling (Emanuel, 2017; I.-I. Lin et al., 2013; Schade & Emanuel, 1999).

We use monthly predictors, including dynamical variables and thermodynamic variables, on a 2.5° grid, rather than hourly or daily fields at finer resolution, to isolate the large-scale environment relevant to PepC-Global while keeping the workflow broadly portable. First, regarding our TC track module, neither hourly nor daily wind fields significantly outperform the use of monthly winds (not shown). Furthermore, previous research has established that monthly data are sufficient for statistical modeling of TC intensity (Lee et al., 2015). Second, this choice reduces contamination of the background state by TCs themselves, since reanalysis can represent observed storms and their associated circulation and moisture anomalies (Hersbach et al., 2020), which can imprint on environmental fields at short time scales and high spatial resolution (Hodges et al., 2017). This matters because PepC-Global is a statistical downscaling framework that conditions on large-scale predictors to generate stochastic TC realizations, not a TC detection or tracking algorithm (Bourdin et al., 2022). Third, coarser, monthly inputs reduce data-handling costs and improve compatibility with climate-model archives, which commonly provide monthly mean fields and relatively coarse horizontal resolution for long historical and scenario integrations (Eyring et al., 2016; Haarsma et al., 2016), whereas explicitly resolving higher-resolution TC-like structure remains computationally expensive and is not uniformly available (Haarsma et al., 2016).

We obtain TC track and intensity information from the International Best Track Archive for Climate Stewardship (IBTrACS), which provides a globally harmonized best-track compilation suitable for model training and evaluation (Knapp et al., 2010). Maintained by NOAA's National Centers for Environmental Information (NCEI), IBTrACS consolidates best tracks from multiple Tropical Cyclone Warning Centers and Regional Specialized Meteorological Centers while retaining agency-reported values to support inter-agency comparison. The archive includes storm identifiers, positions (latitude and longitude), timing in UTC, and intensity metrics such as maximum sustained wind speed and minimum central pressure, along with storm type and related metadata when available (Knapp et al., 2010). IBTrACS spans the historical record from the mid-19th century to the present, with track points mostly reported at 6-hour intervals (00, 06, 12, and 18 UTC) and a 3-hourly product provided via linear interpolation from the 6-hourly records. In this study, we use only the US agency 6-hourly best-track sources within IBTrACS, specifically NHC for NA and the eastern North Pacific (ENP) and JTWC for the other basins.

Figure 1 shows the land-ocean classification at both native (0.25° × 0.25°) and regridded (2.5° × 2.5°) resolutions. A 2.5° grid cell is classified as ocean if more than 50% of its constituent 0.25° cells are over ocean. The figure also displays the seven basin domains (AS, BoB, WNP, ENP, NA, SI, and SP) and the coastlines used for landfall detection. Coastline geometry is derived from the Natural Earth 110m physical vectors dataset.

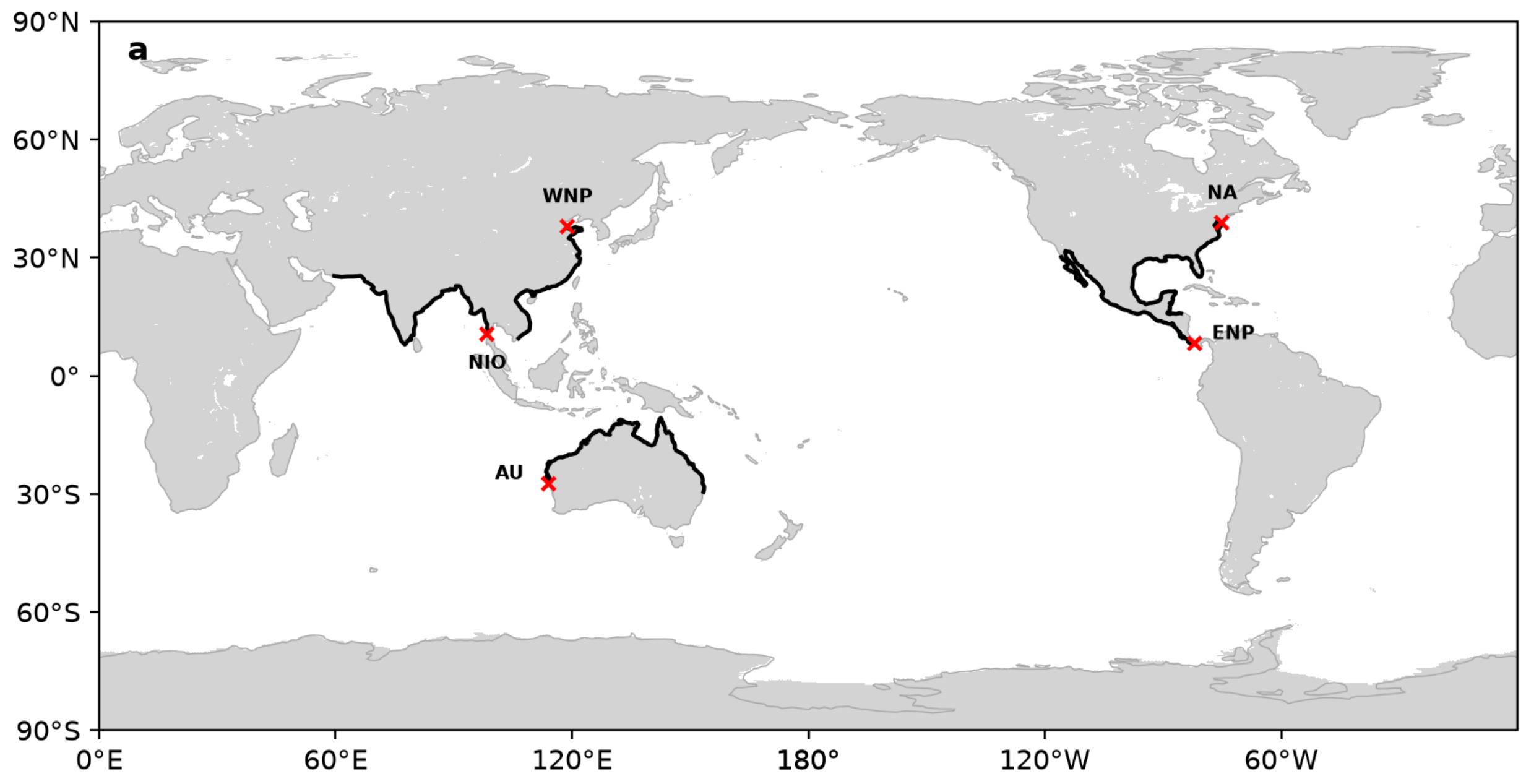

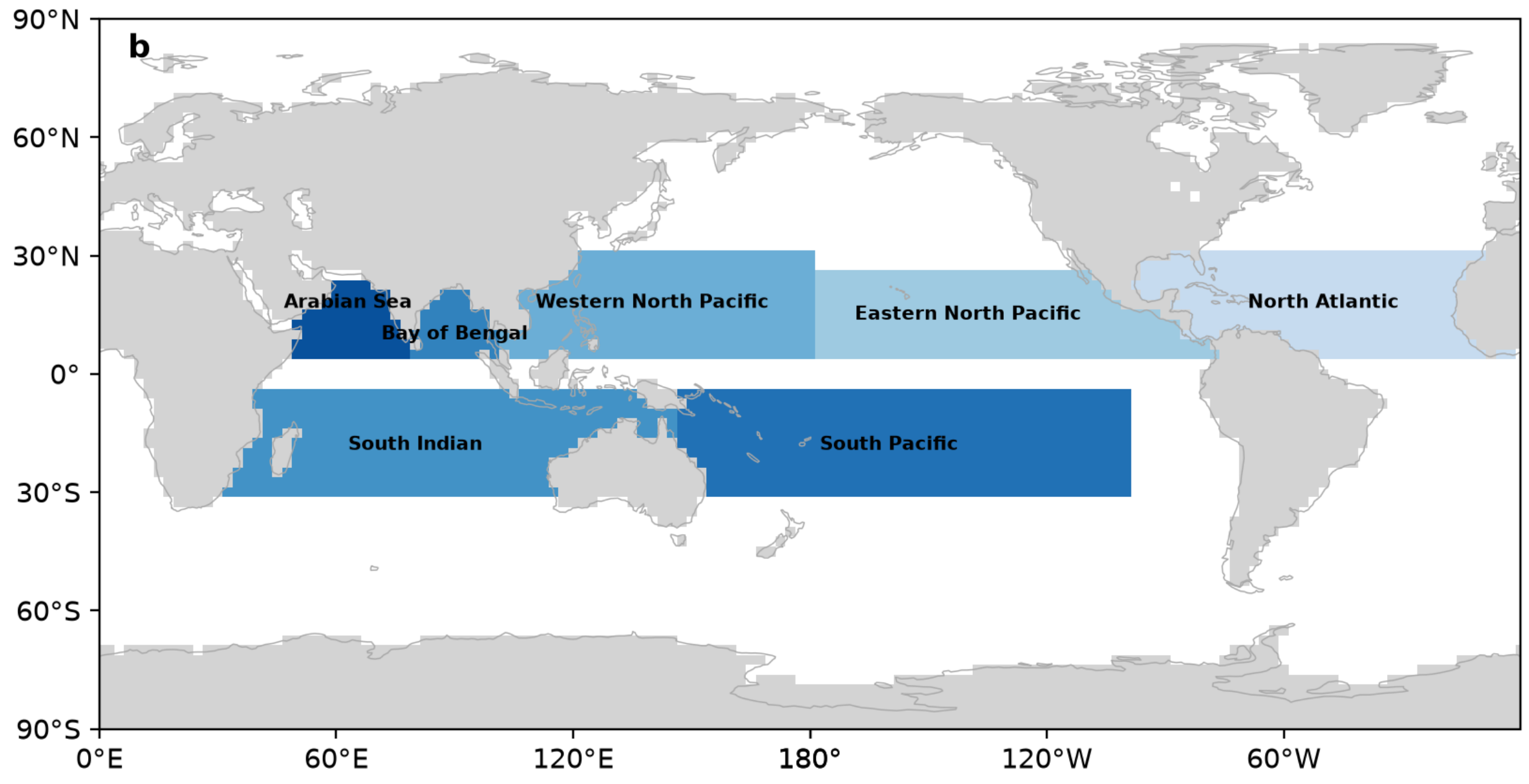

**Figure 1. Basin domains and land-ocean mask.** Land-ocean mask at (a) 0.25° × 0.25° resolution (b) 2.5° × 2.5° resolution. In (a), the five coastlines shown in black, corresponding to the NIO, WNP, NA, ENP, and Australia (AU), are used in the landfall-related evaluation; data are from ERA5. The red X on each coastline marks the starting point for cumulative distance measurement. In (b), each 2.5° × 2.5° grid cell contains 100 0.25° × 0.25° grid cells and is classified as ocean if more than 50 of them are over ocean. The blue shading indicates

the coverage of each basin for TC genesis. If a TC forms within a given basin, it is assigned to that basin for its entire life cycle.

## 2.2 PepC-Global Framework

A schematic overview of the PepC-Global framework is shown in Fig. 2, which summarizes the coupled genesis, track, and intensity modules and their predictor–response structure. The genesis component estimates the dimensionless genesis rate, $\lambda(\text{longitude}, \text{latitude}, \text{time})$, using an SVM classification (SVC) model, with the predictors including 850-hPa absolute vorticity (in units of s$^{-1}$) $\eta_{850} = \zeta_{850} + f$, where $\zeta_{850} = \left(\frac{\partial v}{\partial x} - \frac{\partial u}{\partial y}\right)_{850}$ and $f = 2\Omega sin\varphi$ with Earth's rotation rate $\Omega = 7.292 \times 10^{-5}\ s^{-1}$ and $\varphi$ is latitude, vertical wind shear (in units of m s$^{-1}$) between 850 and 200 hPa defined as $\text{VWS} = [(\text{U850} - \text{U200})^2 + (\text{V850} - \text{V200})^2]^{1/2}$ where U850, U200 and V850, and V200 denote the 850-hPa and 200-hPa zonal and meridional wind components, relative humidity (in units of %) at 600 hPa, and potential intensity (in units of m s$^{-1}$). The track component predicts next 6-hour longitude change (in units of degrees adjusted to its equatorial equivalent) and next 6-hour latitude change (in units of degrees) using a random forest regression model driven by steering-flow predictors U250, V250, U850, and V850, which denote the 250- and 850-hPa zonal (U250, U850) and meridional (V250, V850) wind components (in units of m s$^{-1}$), together with a latitude-dependent term representing the beta-drift effect (cosine of latitude). The intensity component predicts next 6-hour intensity change (in units of knots) using a hidden Markov model with covariates including previous 6-hour intensity change (in units of knots), current intensity (in units of knots), potential intensity (in units of m s$^{-1}$), vertical wind shear (in units of m s$^{-1}$) between 850 hPa and 200 hPa, relative humidity (in units of %) at 600 hPa, and an ocean feedback factor $\text{OCN} = 1 - 0.87 \exp[-0.01 h_m \Gamma^{-0.4} u_T \text{PI} v^{-1}]$, where $h_m$ is ocean mixed-layer depth (in units of m), $\Gamma = T_{\text{seawater}}(z = h_m) - T_{\text{seawater}}(z = h_m + 100)$ is the seawater potential temperature difference (in units of K) between the mixed layer depth and 100 m below it, $u_T$ is TC translation speed (in units of m s$^{-1}$), PI is potential intensity (in units of knots), and $v$ is current TC intensity (in units of knots). Predictors are provided to each module (bottom boxes), and the predicted quantities (top boxes) are used to generate large ensembles of physically plausible synthetic TCs. The intensity module shown is applied only during the over-ocean phase. Over land, an environment-dependent decay model governs intensity evolution.

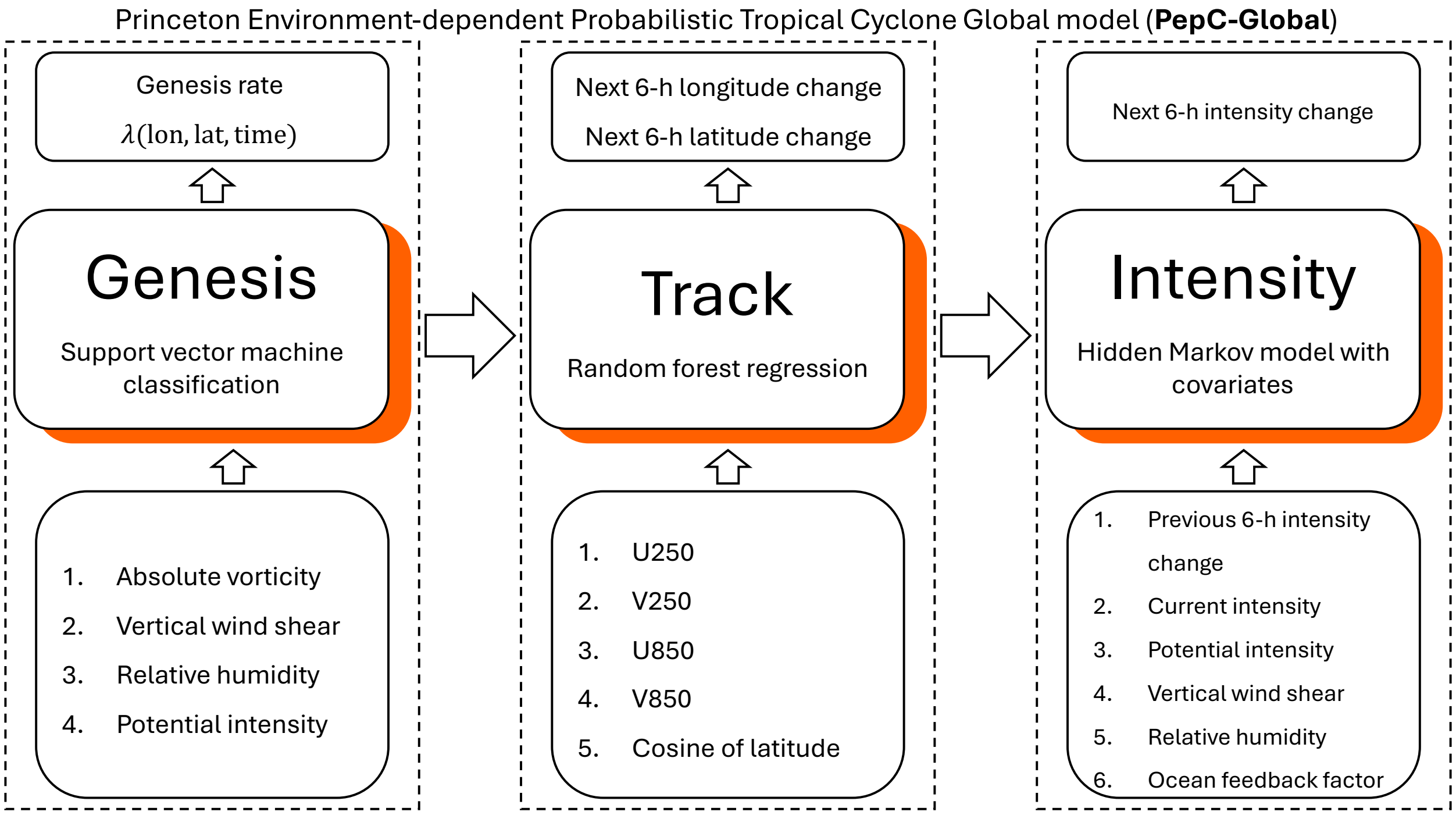

**Figure 2. Schematic of the PepC-Global framework, which links probabilistic models of TC genesis, track, and intensity conditioned on large-scale environmental predictors.**

### 2.3 Experimental design

We conduct a series of numerical experiments to evaluate PepC-Global performance. Figure 3 outlines how observational constraints are applied and relaxed across experiments. Observed genesis events (G0), observed tracks (T0), and observed intensities (I0) provide the reference for evaluation. The genesis module produces simulated genesis rates and simulated genesis events (G1), which provide the initial time and position required to start the track module. Using these genesis inputs, the track module generates simulated tracks beginning from observed genesis time and position (T1 from G0) and simulated tracks beginning from simulated genesis time and position (T2 from G1); because the track module does not define its own termination, these integrations are manually capped at 100 time steps (600 hours), which is longer than 99.9% of observed TC lifetimes from genesis. Given the prescribed track sets, the intensity module produces along-track intensity realizations, including simulated intensities along observed tracks (I1 from T0), along tracks simulated from observed genesis (I2 from T1), and along tracks simulated from simulated genesis (I3 from T2). In the fully coupled life-cycle simulations, track integration is terminated using the intensity evolution, such that a simulated track ends when the simulated intensity falls below 11 knots; the resulting intensity-terminated track collections are denoted T3 (track evolution paired with I1), T4 (track evolution paired with I2), and T5 (track evolution paired with I3).

For genesis, we compare G1 with G0 using basin-wise frequency, seasonal cycle, interannual variability, and spatial distribution. For track, we compare T1 (truncated when the corresponding observation stops) with T0 using basin-wise passage density and along-coastline landfall frequency. For intensity, we compare I1 with I0 using basin-wise lifetime maximum intensity distributions, landfall intensity distributions, and landfall intensity at observed landfall locations. For the fully coupled PepC-Global system, we compare T5 with T0 for track and I3 with I0 for intensity.

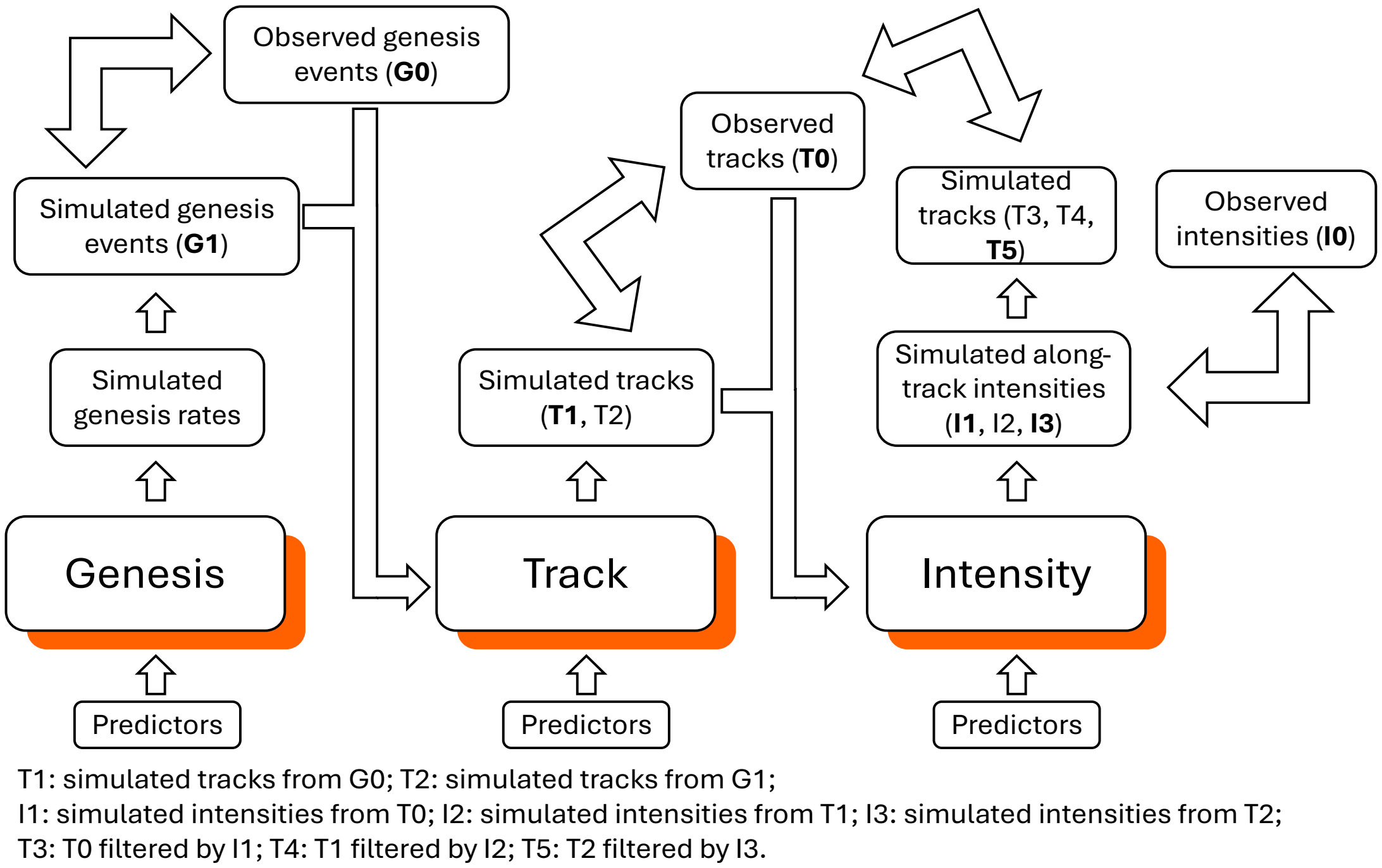

**Figure 3. Schematic of the PepC-Global simulation workflow and the experiment sets.** Comparisons between the bold-labeled quantities (G1 versus G0 for genesis, T1 and T5 versus T0 for track, and I1 and I3 versus I0 for intensity) are presented and discussed in Sections 3–6.

## 3. PepC-Global Genesis Module

The relationship between large-scale environment and TC genesis is inherently probabilistic because favorable large-scale environmental conditions are necessary but not sufficient (Gray, 1968; Sobel et al., 2021; Tippett et al., 2011). Environments characterized by strong low-level (850 hPa) cyclonic vorticity, weak vertical wind shear between 850 hPa and 200 hPa, sufficient midlevel (600 hPa) humidity, and high potential intensity are physically and statistically associated with enhanced genesis likelihood (Jing & Lin, 2020), yet considerable outcome variability remains even under these favorable conditions (Jing & Lin, 2020; Tippett et al., 2011). Under similar large-scale states, mesoscale convective organization and internal vortex dynamics, together with air–sea coupling and storm-induced ocean feedbacks, can modulate the transition from an incipient disturbance (or a seed vortex) to a self-sustaining vortex (Gao et al., 2025). This motivates modeling genesis as a random count process over fixed space–time intervals, which we formalize next using a Poisson framework.

TC frequency (number per fixed spatiotemporal interval; $Y$) is taken as a Poisson random variable, i.e., $Y \sim \mathrm{Poi}(\lambda)$, and the Poisson rate parameter or its expectation is $\lambda$, i.e., $E(Y) = \lambda$. The probability mass function of a Poisson distribution is

$$P(Y = k) = \frac{e^{-\lambda}\lambda^k}{k!}.$$

Obviously, $Y$ at different times and locations does not necessarily follow the same Poisson distribution. Instead, $Y_i \sim \mathrm{Poi}(\lambda_i)$ . And the $\lambda_i$ can be predicted by the environmental predictors $\mathbf{X}$.

The Poisson linear regression assumes a linear relationship between $\log \lambda_i$ and $\mathbf{X}$, with the regression equation given by

$$\log \lambda_i = \mathbf{x}_i \boldsymbol{\beta} = \beta_0 x_{i0} + \beta_1 x_{i1} + \cdots + \beta_p x_{ip}, \qquad x_{i0} \equiv 1$$

or equivalently,

$$\lambda_i = e^{\beta_0 + \beta_1 x_{i1} + \cdots + \beta_p x_{ip}}.$$

However, the log-linear assumption fails to hold when effects of environment on TC genesis are nonlinear. To mitigate this specification risk, we employed SVMs, which replace the fixed linear form with flexible kernel functions that implicitly project inputs into infinite-dimensional feature spaces (Cortes & Vapnik, 1995). SVMs can be used for both classification (finite discrete response) and regression (continuous response). For classification, SVMs learn a function mapping predictors to response by maximizing the margin between classes. For regression, SVMs learn the function by fitting a narrow “tube” around the response. Unlike many models (e.g., deep neural networks) whose training is non-convex and may converge to local minima, SVMs solve a convex optimization with a unique global optimum. SVMs have seen widespread application in climate and atmospheric sciences, including prediction of TC genesis from mesoscale convective systems (T. Zhang et al., 2019), statistical downscaling of precipitation (Tripathi et al., 2006), and various other climate change research applications (Huntingford et al., 2019).

In observations, $y_i$ takes only two values (non-genesis and genesis) in a given month and $2.5° \times 2.5°$ grid cell for more than 99.9% of cases. Thus, the binary outcome from two-class classification models generally encodes $\lambda_i$ . In the binary case $y_i \in \{-1, 1\}$ where −1 represents non-genesis and 1 represents genesis, linear soft-margin SVMs estimate a hyperplane $f(\mathbf{X}) = \mathbf{w}^{\mathrm{T}}\mathbf{X} + b$ by maximizing the geometric margin while allowing limited violations through slack variables $\xi_i$. SVM training involves solving a convex optimization problem, i.e.,

$$\min_{\mathbf{w},b,\xi\geq 0} \frac{1}{2}||w||^2 + C\sum_{i=1}^{n} \xi_i$$

$$\text{s.t.} \;\; y_i(\mathbf{w}^{\mathrm{T}}\mathbf{x}_i + b) \geq 1 - \xi_i, \;\; i = 1, \dots, n,$$

where $C$ trades off margin width against misclassification penalty and is selected based on model performance within the validation dataset. Instead of solving this primal problem directly, it is advantageous to reformulate the problem in its dual form, i.e.,

$$\max_{\alpha} \sum_{i=1}^{n} \alpha_i - \frac{1}{2}\sum_{i=1}^{n}\sum_{j=1}^{n} \alpha_i\alpha_j y_i y_j \mathbf{x}_i^{\mathrm{T}}\mathbf{x}_j$$

$$\text{s.t.} \;\; 0 \leq \alpha_i \leq C, \;\; \sum_{i=1}^{n} \alpha_i y_i = 0.$$

Nonlinear decision boundaries arise via the kernel trick, replacing inner products $\mathbf{x}_i^{\mathrm{T}}\mathbf{x}_j$ with a kernel $k(\mathbf{x}_i, \mathbf{x}_j)$.

$$\max_{\alpha} \sum_{i=1}^{n} \alpha_i - \frac{1}{2}\sum_{i=1}^{n}\sum_{j=1}^{n} \alpha_i\alpha_j y_i y_j k(\mathbf{x}_i, \mathbf{x}_j)$$

$$\text{s.t.} \;\; 0 \leq \alpha_i \leq C, \;\; \sum_{i=1}^{n} \alpha_i y_i = 0,$$

and predictions rely only on the SVs:

$$\hat{y}(\mathbf{X}) = \mathrm{sign}\left(\sum_{i \in \mathrm{SV}} \alpha_i y_i k(\mathbf{x}_i, \mathbf{X}) + b\right),$$

where $\mathbf{x}_i$ is one of the SVs. Here we use the radial basis function, also known as the Gaussian kernel,

$$k_{\mathrm{Gaussian}}(\mathbf{x}_i, \mathbf{X}) = e^{-\gamma||\mathbf{x}_i - \mathbf{X}||^2},$$

which induces an infinite-dimensional reproducing kernel Hilbert space. Crucially, SVMs with the Gaussian kernel can represent any continuous function given sufficient data and appropriate regularization, while retaining convex optimization and margin-based generalization guarantees.

Most TC genesis potential indices relate large-scale environmental predictors to genesis frequency using the Poisson linear regression framework (Sobel et al., 2021; Tippett et al., 2011) including the PepC genesis module (Jing & Lin, 2020), or closely related regression formulations (Emanuel & Nolan, 2004; Murakami & Wang, 2010; B. Wang & Murakami, 2020; M. Zhang et al., 2016). Here, the PepC-Global genesis module instead casts genesis as a classification problem solved with SVM classification using a Gaussian kernel. We treat the SVM prediction as the Poisson rate parameter, from which TC genesis events are sampled. Consistent with the PepC genesis module, we retain absolute vorticity and vertical wind shear as two dynamic predictors, and relative humidity and potential intensity as two thermodynamic predictors. Using the same predictors, SVC significantly outperforms Poisson linear regression across basins. Note that the clustering algorithm introduced in the PepC genesis module is not retained in the PepC-Global genesis module. That algorithm was used to reconcile the spatial continuity of environmental fields with the discreteness of storm occurrences, but this is no longer necessary with SVM.

Table 1 summarizes spatial skill in reproducing 40-year genesis frequency at the grid-cell scale within each basin, measured by the correlation coefficient between observed and simulated frequency (prior to Poisson sampling). The Poisson linear regression exhibits near-zero skill in AS and NA and comparatively low skill in BoB, while correlations are higher in WNP, ENP, and SI and highest in SP, which might indicate pronounced basin-to-basin differences in grid-scale predictability. Note that the PepC genesis module applies a clustering algorithm prior to Poisson linear regression (Jing & Lin, 2020) and achieves better results than the standard Poisson linear regression shown here. In contrast, SVC achieves consistently strong training correlations and improves test performance in all basins except SP, with particularly large gains in NA and ENP, although skill remains more modest in WNP and SI.

**Table 1. Skill of the PepC-Global genesis module in reproducing climatological genesis frequency at the grid-cell scale**. Skill is measured by the correlation coefficient between observed and simulated frequency across grid cells within each basin. Results are shown for a Poisson linear regression ("Linear") and the SVC model, each evaluated on the same 80%/20% train/test split.

| | AS | BoB | WNP | ENP | NA | SI | SP |
|---|---|---|---|---|---|---|---|
| Linear train | 0.21 | 0.39 | 0.64 | 0.65 | 0.35 | 0.66 | 0.73 |
| Linear test | 0.06 | 0.18 | 0.64 | 0.17 | 0.07 | 0.47 | 0.71 |
| SVC train | 0.99 | 0.99 | > 0.99 | 0.97 | 0.97 | 0.99 | 0.89 |
| SVC test | 0.30 | 0.57 | 0.74 | 0.66 | 0.42 | 0.63 | 0.63 |

The PepC-Global genesis module reproduces the basin-dependent seasonal cycle of TC genesis based on 100 realizations, capturing both the timing and the relative amplitude of peak activity across the seven basins (Fig. 4). This behavior contrasts with a common limitation of existing statistical genesis models, which often produce seasonal cycles that are smoother than observed because Poisson linear regression formulations tend to assign a small but nonzero genesis rate even in months when observed genesis is essentially absent (Sobel et al., 2021; Tippett et al., 2011; Yang et al., 2021). The SVC formulation can represent threshold-like decision boundaries, so the PepC-Global genesis module can assign effectively zero genesis rate parameter during suppressed seasons while still capturing sharp transitions into active periods. Consistent with this, Fig. 4 shows that the PepC-Global genesis module reproduces the bimodal seasonality in NIO, with peaks in late spring to early summer and again in autumn for AS and BoB, as well as the late-summer maximum in WNP, ENP, and NA and the austral-summer peak in SI and SP. Figure 4h further indicates closer agreement between predicted and observed monthly frequency for the PepC-Global genesis module (SVC-based Poisson process) than for Poisson log-linear regression, with smaller root mean square error across basins except SP.

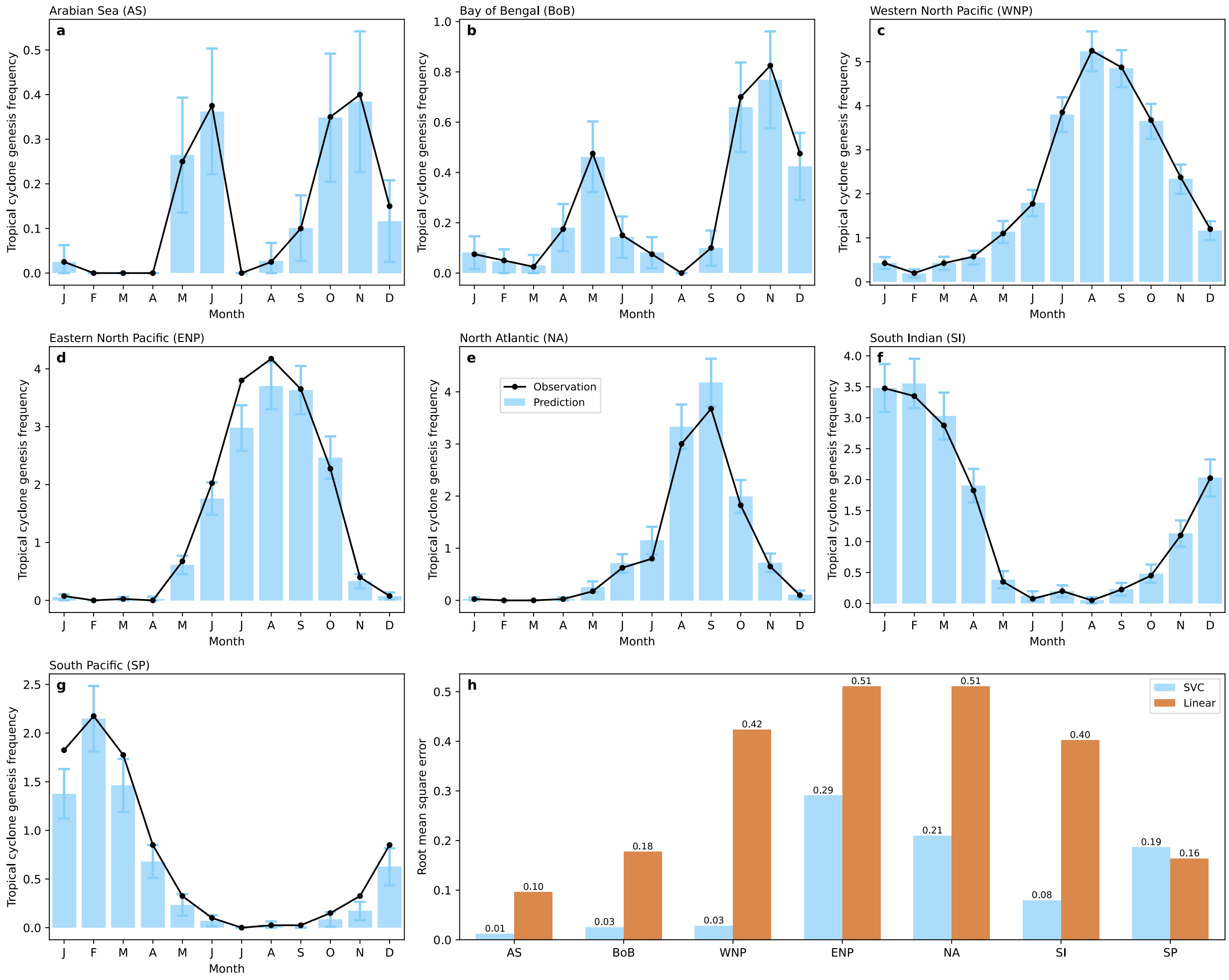

**Figure 4. Basin-wise evaluation of seasonal cycle of TC genesis frequency simulated by the PepC-Global genesis module.** In (a–g), the black lines with markers show observed

mean monthly TC genesis frequency, and blue bars show mean predictions from 100 realizations of the PepC-Global genesis module. Blue error bars indicate plus or minus one standard deviation. In (h), blue and orange bars show the root mean square error for the mean predictions from 100 realizations of the PepC-Global genesis module (SVC-based Poisson process) and Poisson linear regression, respectively.

The PepC-Global genesis module captures much of the interannual variability in basin-wide TC genesis frequency from 1980 to 2019 (Fig. 5). This level of basin-wise agreement is notably stronger than what is typically achieved by existing genesis indices, whose interannual performance is often limited by an inability to reproduce the observed amplitude of year-to-year cyclogenesis fluctuations (Menkes et al., 2011). Consistent with this broader picture, B. Wang and Murakami (2020) reported only modest correlations between their dynamic genesis potential index and basin-total TC genesis in several basins, particularly in NIO, ENP and SI, even after improvement relative to earlier formulations. In contrast, annual frequency predicted by PepC-Global genesis module (SVC-based Poisson process) tracks observed year-to-year changes across basins, with substantially higher correlation coefficients in BoB, WNP, SI and SP, and significant improvement in AS, ENP, and NA compared to Poisson linear regression (Fig. 5). In SP, although the PepC-Global genesis module does not show improved skill in seasonal variability (Fig. 4h), it achieves much higher performance in annual variability (Fig. 5h), compared with Poisson linear regression.

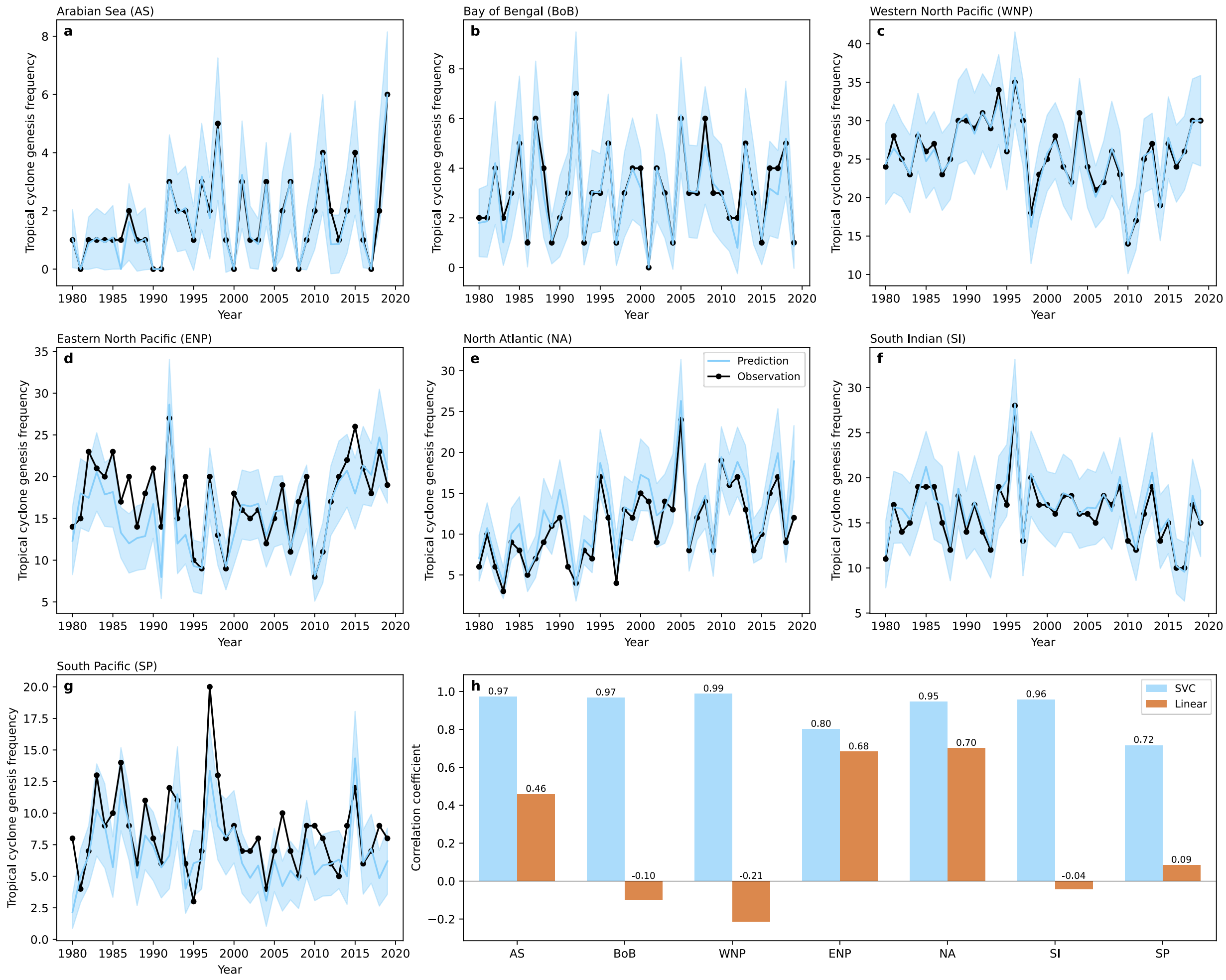

**Figure 5. Basin-wise evaluation of interannual variability of TC genesis frequency during 1980–2019 simulated by the PepC-Global genesis module.** In (a–g), black lines with markers show observed annual TC genesis frequency, and blue lines with markers show mean predictions from 100 realizations of the PepC-Global genesis module. Blue shading indicates plus or minus one standard deviation. In (h), blue and orange bars show the correlation coefficient for mean predictions from 100 realizations of the PepC-Global genesis module (SVC-based Poisson process) and Poisson linear regression, respectively.

The PepC-Global genesis module largely avoids the underprediction in peak-season genesis hotspots that has been documented for smooth Poisson linear regression genesis indices, including underestimation in the tropical Atlantic main development region in earlier index-based studies and also in the PepC genesis configuration (Jing & Lin, 2020; Tippett et al., 2011). We evaluate this behavior using spatial distribution maps (Fig. 6): for each grid cell, the observed genesis frequency is compared with the distribution from 100 PepC-Global realizations, and grid cells are labeled accurate, overestimated, or underestimated depending on whether the observation falls within the 2.5th to 97.5th percentile interval of the 100-member ensemble. We also assess spatial pattern agreement by computing the basin-wise pattern correlation coefficient between observations and the mean grid-cell genesis frequency across the 100 realizations (Fig. 6h). The PepC-Global genesis module (blue bars) shows significantly higher correlations than Poisson log-linear regression (orange bars) across basins, especially for AS, BoB, and NA.

The spatial hotspot underprediction and the "too smooth" seasonal and interannual variability (Menkes et al., 2011) share a common origin: Poisson log-linear regression spreads probability mass too broadly. As a result, it assigns nonnegligible genesis rates during suppressed seasons when observed genesis is essentially absent, while failing to concentrate enough probability in peak regions and periods to match observed amplitudes (Menkes et al., 2011; Sobel et al., 2021; Tippett et al., 2011; Yang et al., 2021). The PepC-Global genesis module overcomes this limitation by employing an SVC-based Poisson process.

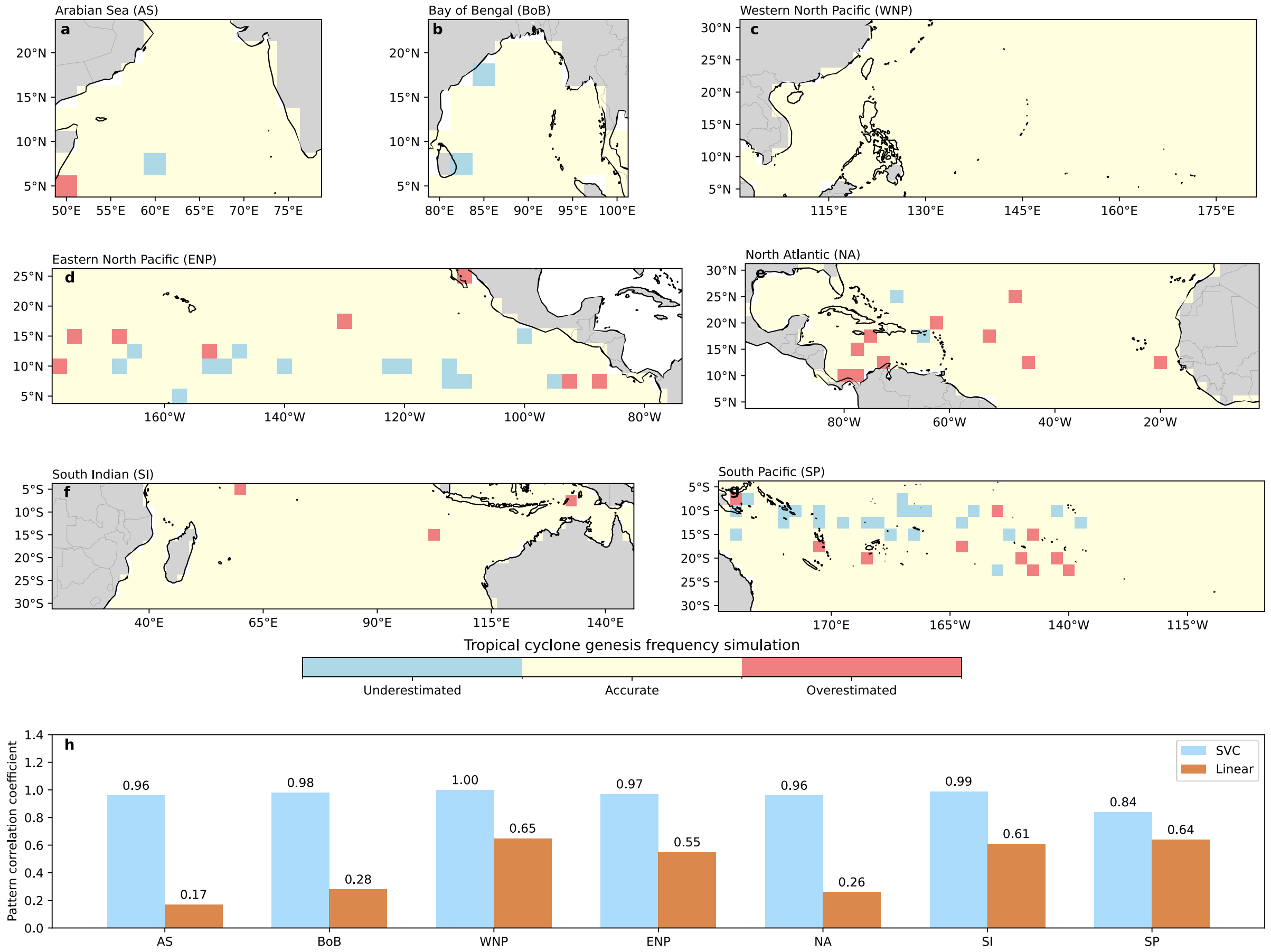

**Figure 6. Basin-wise evaluation of spatial distribution of TC genesis frequency simulated by the PepC-Global genesis module.** In (a–g), climatological TC genesis frequency is compared between observations and 100-realization ensembles from the PepC-Global genesis module for each basin: grid cells are classified as accurate (yellow) when the observation lies between the 2.5th and 97.5th percentiles of the 100-member ensemble, overestimated (red) when the observation falls below the 2.5th percentile, and underestimated (blue) when the observation exceeds the 97.5th percentile. In (h), blue and orange bars show the pattern correlation coefficient for mean predictions from 100 realizations of the PepC-Global genesis module (SVC-based Poisson process) and Poisson linear regression, respectively.

## 4. PepC-Global Track Module

TC translation defines the geographic reach of the threat, as a storm steered into an unfavorable environment or away from vulnerable populations is effectively neutralized and can do no harm. One widely used track model is the BAM, originally developed by Holland (1983) and later adapted for operational use by Marks (1992). The BAM formulation is

$$\mathbf{V}_{\mathrm{TC}} = \mathbf{V}_{\mathrm{steer}} + \mathbf{V}_{\mathrm{beta}},$$

where $\mathbf{V}_{\mathrm{TC}}$ is TC translation velocity, $\mathbf{V}_{\mathrm{steer}}$ is the environmental steering flow and $\mathbf{V}_{\mathrm{beta}}$ is the beta-drift correction. BAM served as part of the US National Hurricane Center guidance suite for decades, with three versions in operational use. The versions differ in the depth over which the steering flow is calculated: the shallow version uses winds averaged over 850–700 hPa, the medium version over 850–400 hPa, and the deep version over 850–200 hPa. The BAM model was also used in KE08 and CHAZ. The formulation of $\mathbf{V}_{\mathrm{steer}}$ in the CHAZ track module is

$$\mathbf{V}_{\mathrm{steer}} = \alpha\mathbf{V}_{850} + (1-\alpha)\mathbf{V}_{250},$$

where α is set to 0.8. The formulation is close to the BAM deep version.

Following the same idea, the PepC-Global track module learns TC motion as an environment-conditioned displacement process via random forest regression (also used in the PepC track module), predicting changes in longitude and latitude. Random forests combine many decision trees fit to bootstrap-resampled data and decorrelate the trees by randomly subsetting predictors at each split, which helps capture nonlinear responses and predictor interactions while limiting overfitting. In our track framework, separate regressors are trained for the zonal and meridional components of storm displacement over a fixed 6-hour step, with displacements defined in latitude and longitude degrees. Predictors include large-scale steering flow information from winds at 250 hPa and 850 hPa, consistent with BAM and the PepC track module (Jing & Lin, 2020), together with a latitude-dependent term to account for beta-drift corrections (Holland, 1983) that is not included in the PepC track module. Zonal displacements are adjusted by the cosine of latitude so that the PepC-Global track module learns a comparable physical displacement scale across latitudes. The resulting ensemble-mean prediction yields the expected motion. The PepC-Global track module builds on the random forest regression, transitioning from an analog-dependent formulation to a wind-only motion model. In the PepC track module, random forest regression is used to map large-scale predictors to 6-hour displacements, but it also relies on historical-track analogs that can implicitly encode basin-specific structure. The PepC-Global track module removes this explicit dependence on historical tracks and instead predicts motion directly from the large-scale steering flow, which makes the track model less tied to the availability of close analogs and more directly linked to physically interpretable controls on storm motion. To benchmark the PepC-Global track module against BAM, we develop a multiple linear regression model based on the BAM formulation:

$$\mathrm{U_{TC}} = a_0 + a_1\mathrm{U_{850}} + a_2\mathrm{U_{250}} + a_3\mathrm{V_{850}} + a_4\mathrm{V_{250}} + a_5\cos\theta,$$

$$\mathrm{V_{TC}} = b_0 + b_1\mathrm{U_{850}} + b_2\mathrm{U_{250}} + b_3\mathrm{V_{850}} + b_4\mathrm{V_{250}} + b_5\cos\theta.$$

The track models used in KE08 and CHAZ are special cases of the BAM-based linear model. Consequently, the BAM-based linear model is expected to perform at least as well as those in KE08 and CHAZ. Note that the PepC-Global track module and the BAM-based linear model share the same predictors. The four wind predictors are horizontally averaged winds over the surrounding 25 grid cells (2.5° × 2.5°), with the 5 nearest cells excluded, to minimize influence from the TC's own circulation.

The skill of the BAM-based linear model and the PepC-Global track module in predicting 6-hour TC displacements in longitude and latitude is quantified across basins using correlation coefficients, with the same 80%/20% train/test split. Table 2 reports results for the one-step-ahead zonal displacement after converting longitude change to an equator-equivalent distance, and Table 3 reports results for the one-step-ahead meridional displacement. As shown in Tables 2 and 3, the linear model exhibits modest skill, with correlations as low as 0.23 for meridional motion. The PepC-Global track module achieves substantially higher correlations, exceeding 0.68 for both zonal and meridional displacements.

**Table 2. Skill in predicting the one-step-ahead (6-hour) longitude change.** Longitude change is adjusted to its equatorial equivalent. Skill is quantified by the correlation coefficient between observed and predicted values. Results are shown for the BAM-based linear model ("Linear") and the PepC-Global track module ("RF"), each evaluated on the same 80%/20% train/test split.

| | AS | BoB | WNP | ENP | NA | SI | SP |
|---|---|---|---|---|---|---|---|
| Linear train | 0.38 | 0.58 | 0.70 | 0.56 | 0.78 | 0.46 | 0.45 |
| Linear test | 0.41 | 0.53 | 0.71 | 0.55 | 0.77 | 0.46 | 0.46 |
| RF train | 0.98 | 0.97 | 0.98 | 0.98 | 0.98 | 0.98 | 0.98 |
| RF test | 0.85 | 0.74 | 0.84 | 0.81 | 0.87 | 0.78 | 0.76 |

**Table 3. Same as Table 2 but for predicting the latitude change.**

| | AS | BoB | WNP | ENP | NA | SI | SP |
|---|---|---|---|---|---|---|---|
| Linear train | 0.23 | 0.45 | 0.54 | 0.40 | 0.41 | 0.43 | 0.47 |
| Linear test | 0.28 | 0.42 | 0.54 | 0.39 | 0.42 | 0.41 | 0.48 |
| RF train | 0.96 | 0.97 | 0.97 | 0.97 | 0.97 | 0.97 | 0.97 |
| RF test | 0.73 | 0.74 | 0.72 | 0.69 | 0.68 | 0.71 | 0.71 |

Neither KE08 nor CHAZ uses monthly winds directly to drive the track module. Instead, both generate synthetic winds as a continuous function of time using Fourier series, which are then used to propagate TCs. However, it is unclear whether synthetic winds compromise physical consistency with other environmental variables if synthetic fields are not also generated for those variables. For consistency, synthetic winds should also be used in the genesis and intensity modules. Additionally, the stochastic nature of synthetic winds complicates model training, which may explain why the BAM coefficients are set somewhat subjectively rather than fitted from data in KE08 and CHAZ. Whether this degree of technical difficulty and theoretical complexity is warranted remains an open question. Here, we add Gaussian noise to the PepC-Global track module predictions to account for both synoptic wind variability and the influence of the TC itself on the relationship between TC motion and large-scale steering flow. Based on the root-mean-square error of one-step-ahead predictions, we choose Gaussian noise with a standard deviation of 0.5°. Sensitivity tests with nearby values yield similar results.

Using Gaussian noise, we drive both the PepC-Global track module and the BAM-based linear model to generate full tracks beginning from observed genesis times and locations. We do not couple the genesis module here to avoid errors it might introduce. Both models are run for 100 realizations. Because the track models terminate after 100 steps or upon reaching polar regions, simulated lifetimes can be much longer than observed, leading to overestimation of track passages. To avoid this, we truncate each simulated track based on the observed lifetime; that is, simulated and observed TCs with the same genesis time and location are assigned the same lifetime. We do not couple the intensity module here to avoid errors it might introduce. However, we note that this truncation method may also introduce some bias, as lifetime is not fully determined by genesis time and location alone.

After truncating the tracks, the spatial distribution of simulated TC passages from the PepC-Global track module is compared with observations in Fig. 7. For each 2.5° × 2.5° grid cell, we compare the observed passage count with the ensemble distribution of passage counts from the 100 realizations. Note that the evaluation includes both ocean and land areas. For most regions, the PepC-Global track module reproduces the observed distribution of TC passages. However, significant biases are found in WNP: the model underestimates passages over the South China Sea while overestimating passages over the ocean near 45°N. A possible explanation is that fewer TCs in the model recurve toward South Asia and the South China Sea, while more propagate toward Japan and South Korea. This discrepancy may arise from the complex land-sea distribution associated with marginal seas and islands affecting the representation of large-scale circulation. We also compute the pattern correlation coefficient between observed grid-cell passage counts and the ensemble-mean passage counts, which is high (at least 0.95) across all basins. Compared to the BAM-based linear model, the PepC-Global track module shows better performance. The close agreement between observations and simulations suggests that the simple Gaussian noise method may serve as an effective alternative to synthetic wind generation.

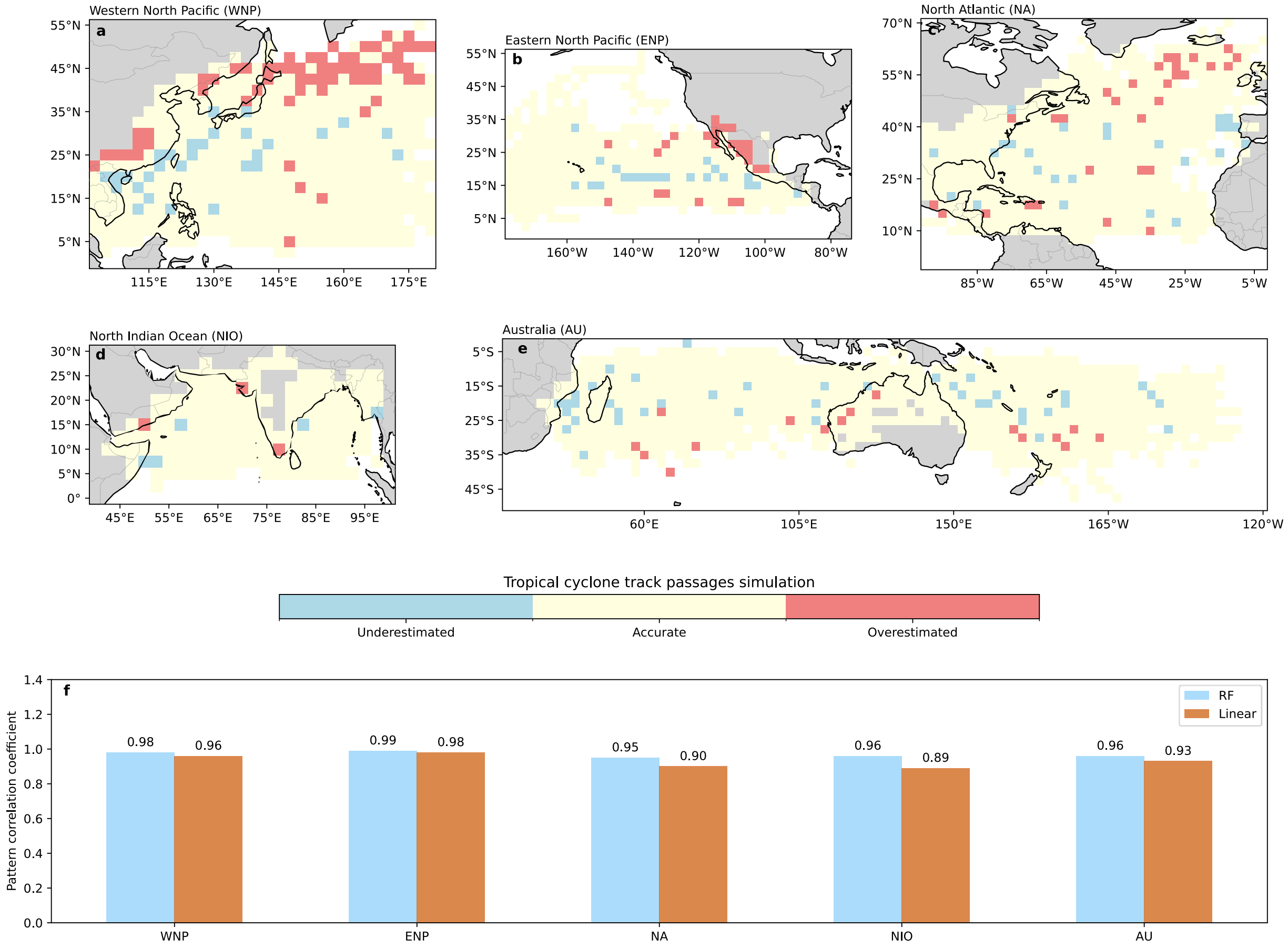

**Figure 7. Basin-wise evaluation of spatial distribution of TC track passages simulated by the PepC-Global track module for simulations initialized from observed genesis.** The comparison is restricted to grid cells with at least one observed TC passage. In (a–e), climatological TC passage counts per grid cell are compared between observations and 100-realization ensembles from the PepC-Global track module for each basin: grid cells are classified as accurate (yellow) when the observation lies between the 2.5th and 97.5th percentiles of the 100-member ensemble, overestimated (red) when the observation falls below the 2.5th percentile, and underestimated (blue) when the observation exceeds the 97.5th percentile; white areas indicate grid cells with no observed passages. In (f), blue and

orange bars show the basin-wise spatial pattern correlation coefficients between observed and ensemble-mean simulated passage counts for the PepC-Global track module (random forest regression) and the BAM-based linear model, respectively. Both models include Gaussian noise with a standard deviation of 0.5°.

Given that landfall is the primary source of TC-induced losses, the ability to reproduce observed landfall distributions is a crucial test of track model skill. TC tracks and coastlines are each represented as connected line segments. We identify landfall by computing all pairwise segment intersections and selecting the first occurrence along the track. As Fig. 8 shows, the PepC-Global track module reproduces along-coastline landfall distributions very well. The spread across 100 realizations of the PepC-Global track module is comparable to the observational uncertainty estimated via bootstrap resampling. Furthermore, the ensemble mean agrees closely with observations, as indicated by the low Hellinger distances.

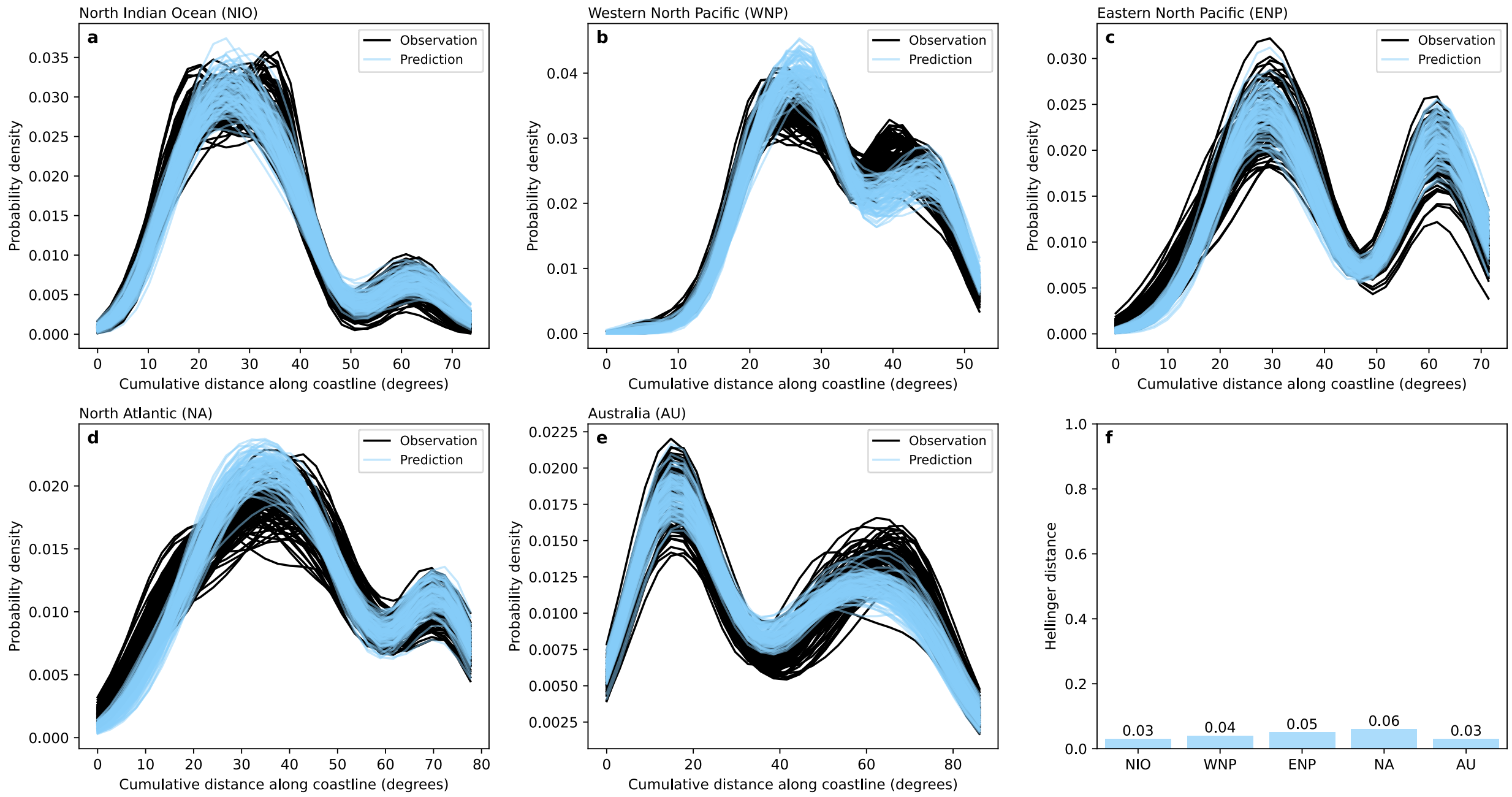

**Figure 8. Basin-wise evaluation of spatial distribution of along-coastline distributions of TC landfall frequency simulated by the PepC-Global track module for simulations**

**initialized from observed genesis.** In (a–e), probability density of landfall positions is shown as a function of cumulative distance (in units of degrees) along the coastline for each basin, with observations in black and the PepC-Global track module simulations in blue. Observational uncertainty is represented by 100 curves (the observed sample and 99 bootstrap resamples); simulations comprise 100 realizations. In (f), blue bars show the Hellinger distance between the observed and the mean simulated distributions for each basin (smaller values indicate closer agreement).

## 5. PepC-Global Intensity Module

Statistical intensity guidance remains a cornerstone of operational TC forecasting because it provides fast, well-calibrated benchmarks for predictable intensity change. One widely used example is the Statistical Hurricane Intensity Prediction Scheme (SHIPS), which applies multiple linear regression to relate intensity change to storm and environmental predictors (DeMaria et al., 2005; DeMaria & Kaplan, 1994, 1999). By fitting these empirical relationships to large historical samples, SHIPS is computationally efficient and straightforward to implement, while its regression structure offers a transparent summary of how predictors contribute to expected intensity evolution. Despite being far simpler than full-physics numerical weather prediction, SHIPS has long served as a high-skill baseline and has remained competitive for some lead times and regimes when compared against operational benchmarks (Cangialosi et al., 2020). CHAZ adopts a similar approach, modeling intensity change as a linear regression on environmental and storm predictors with additive white noise to represent stochastic variability (Lee et al., 2016a, 2018).

The Markov Environment-Dependent Hurricane Intensity Model (MeHiM) substantially improves upon SHIPS in capturing TC intensity evolution (Jing & Lin, 2019). MeHiM offers two key advantages over SHIPS. First, MeHiM predicts TC intensity change based on both environmental variables and hidden states. Here, the hidden states might represent the internal properties of the TC, such as vortex structure, eyewall organization, and warm core strength. These internal properties govern how sensitively the TC responds to its surrounding environment. Second, MeHiM models the temporal evolution of states as a Markov chain. Since the states can reflect TC internal properties, they are not independent between consecutive time steps. Because MeHiM uses only large-scale environmental predictors as input, the states are hidden. In summary, MeHiM is a hidden Markov model with covariates (environmental predictors), also known as an input-output hidden Markov model (Bengio & Frasconi, 1996), which maps input sequences (environmental predictors) to output sequences (intensity change sequences). It models intensity change as a discrete dynamic system defined by

$$y_t = g(q_t, \mathbf{X}_t),$$

$$q_t = f(q_{t-1}, \mathbf{X}_t),$$

where $\mathbf{X}_t$ is the input vector at time $t$, $y_t$ is the output scalar at time $t$, and $q_t$ is the hidden state at time $t$. In MeHiM, $\mathbf{X}$ includes the prior 6-hour intensity change, current intensity, potential intensity, vertical wind shear between 850 hPa and 200 hPa, relative humidity at 600 hPa, and an ocean feedback factor OCN.

The PepC-Global intensity module couples MeHiM (over-ocean intensity component) with an over-land intensity model, switching between them based on whether the storm remains over ocean during the next 6-hour step. MeHiM is applied only when both the current and the next 6-hour positions are over ocean; any land-interacting segment (ocean to land, land to ocean, or land to land) is handled by the over-land model. Using the ERA5 land-ocean mask at 0.25° × 0.25° resolution, each 2.5° × 2.5° grid cell is classified as ocean if its ocean fraction exceeds 0.5, and as land otherwise. The labels of 2.5° × 2.5° grid cells are then used to determine whether a TC is over ocean or land at each 6-hour step.

We upgrade the over-land intensity model from the fixed decay rate model used in Jing and Lin (2019) to an environment-dependent formulation. The original model follows the widely used post-landfall decay framework of Kaplan and DeMaria (1995):

$$v + \delta v = v_b + (v - v_b)e^{-\alpha \delta t},$$

where $v$ is current intensity, $\delta v$ is next 6-hour intensity change, $v_b$ is the intensity that a TC may maintain over land (set to 10 knots), $\alpha$ is a constant that does not vary with environmental conditions, $\delta t$ is 6 hours.

Motivated by the importance of narrow landmasses (DeMaria et al., 2006) and the strong sensitivity of decay to the surrounding environment (Wong et al., 2008), the upgraded formulation predicts 6-hour intensity change using environmental predictors and ocean fraction at both the current and next positions, allowing decay rates to vary across storms and locations rather than being prescribed as constant. Following Chen and Chavas (2020, 2021), who showed that surface drying is a key mechanism governing the transient intensity response of tropical cyclones upon landfall, we include soil moisture and soil temperature as over-land predictors. The environment-dependent decay model takes the form,

$$v + \delta v = v_b + (v - v_b)\mathrm{e}^{-\alpha \bar{L}},$$

$$\alpha = \exp(\beta_0 + \beta_1 E),$$

$$\bar{L} = 0.2(1 - \mathrm{Ocean}) + 0.8(1 - \mathrm{Ocean}_{\mathrm{next}}),$$

$$\bar{\theta}_s = \frac{0.2\theta_s(1 - \mathrm{Ocean}) + 0.8\theta_{s,\mathrm{next}}(1 - \mathrm{Ocean}_{\mathrm{next}})}{\bar{L}},$$

$$\overline{T_s} = \frac{0.2T_s(1 - \mathrm{Ocean}) + 0.8T_{s,\mathrm{next}}(1 - \mathrm{Ocean}_{\mathrm{next}})}{\bar{L}},$$

$$E = \bar{\theta}_s \max(0, \exp[0.07(\overline{T_s} - T_c)] - 1),$$

where $\theta_s$ is the volumetric soil water content (0–7 cm depth; in units of $m^3$ $m^{-3}$) at the current position, $T_s$ is the soil temperature (0–7 cm depth; in units of K) at the current position, Ocean is the ocean fraction (dimensionless) at the current position, $\theta_{s,\mathrm{next}}$, $T_{s,\mathrm{next}}$, and $\mathrm{Ocean}_{\mathrm{next}}$ denote the corresponding values at the next 6-hour position. $\bar{L}$ is the land exposure of the 6-hour step; $\bar{\theta}_s$ and $\overline{T_s}$ are the land-weighted means of soil water content and soil temperature, respectively; $E$ is a surface flux proxy, with $T_c = 299.0$ K a fixed global constant; $E$ enters in raw units, so that $E = 0$ indicates no enthalpy support from land; and $\alpha$ is the decay rate per 6-hour step. The weights of 0.8 and 0.2 are determined from sensitivity tests, which indicate that intensity change over the next 6 hours is more sensitive to land surface conditions at the destination than at the departure point. For each basin, we estimate the fixed decay rate and the environment-dependent model parameters from the training dataset and apply the trained models to the test dataset. We repeat this procedure for 1,000 random 80%/20% train/test splits. As shown in Table 4, the environment-dependent over-land decay model achieves a lower test-set RMSE than the fixed decay rate model in at least 799 of the 1,000 splits in every basin, including all 1,000 splits in SI.

**Table 4. Number of wins, out of 1,000 random train/test splits, of** the environment-dependent over-land decay model ("Env") over the fixed decay rate model ("Fix") in predicting one-step-ahead (6-hour) over-land intensity change. A win means that Env has a lower root-mean-square error (RMSE) than Fix on the test set of a given 80%/20% train/test split.

| | AS | BoB | WNP | ENP | NA | SI | SP |
|---|---|---|---|---|---|---|---|
| Env wins | 799 | 998 | 995 | 952 | 980 | 1000 | 922 |

We evaluate the coupled intensity module in terms of modeled distributions of lifetime maximum intensity, RI, and landfall intensity. Because lifetime maximum intensity largely determines a storm's damage potential, accurate reproduction of lifetime maximum intensity statistics is an important test for TC intensity models. From a climatological perspective, the lifetime maximum intensity distribution is a fundamental feature of the climate system. The distribution of lifetime maximum intensity does not follow the typical pattern of extreme events, in which more intense events are progressively rarer. Instead, the lifetime maximum intensity distribution exhibits a secondary peak at high intensities, indicating that the most extreme storms can be more frequent than moderately intense ones (Lee et al., 2016b). Note that this second peak feature is more pronounced in WNP, ENP and SI, as shown in Fig. 9. This distinctive feature has challenged previous TC intensity models, such as the one used in CHAZ (Lee et al., 2016a). The blue curves in Fig. 9 show the lifetime maximum intensity distributions reproduced by 100 realizations of the PepC-Global intensity module using observed tracks. Compared with the observations and 99 bootstrap resamples (black curves), the predictions show high consistency. Moreover, the spread of the blue curves is comparable to that of the black curves. The Hellinger distances between the observed and ensemble-mean predicted distributions are very low, indicating close agreement. In summary, given observed tracks and driven by along-track environmental predictors, the PepC-Global intensity module reproduces the observed lifetime maximum intensity distribution with high fidelity and spread comparable to bootstrap resampling uncertainty. Jing and Lin (2019) demonstrated that MeHiM outperforms linear models in NA. Following their approach, we verify whether MeHiM also outperforms a linear model with the same predictors in other basins. As shown in Fig. 9h, MeHiM achieves higher agreement with observations than the linear model, particularly for WNP, ENP, and SI, where the bimodal structure in lifetime maximum intensity distributions is most pronounced and linear models struggle to reproduce the second peak.

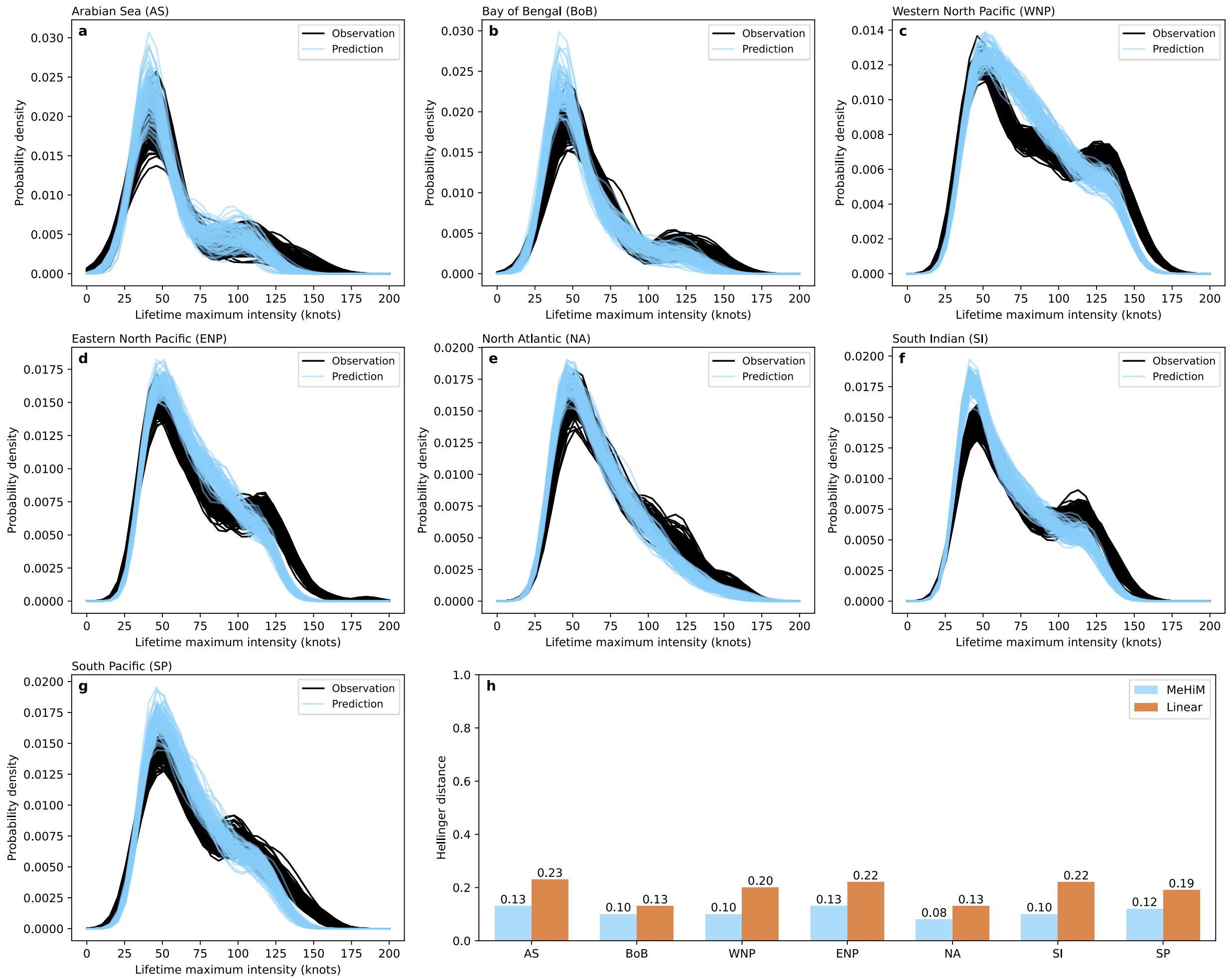

**Figure 9. Basin-wise evaluation of lifetime maximum intensity distributions simulated by the PepC-Global intensity module along observed tracks.** In (a–g), blue curves show probability densities of lifetime maximum intensity computed from 100 PepC-Global intensity realizations generated along observed tracks. Black curves show the probability densities of the observed lifetime maximum intensity and its 99 bootstrap resamples (100 curves total). In (h), blue and orange bars show the Hellinger distance between the observed probability density and the median simulated probability density for the PepC-Global over-ocean intensity module (MeHiM) and the linear model, respectively. Both models are coupled with the same environment-dependent over-land decay model and include the same Gaussian noise.

Accurate prediction of RI is important not only for weather forecasting but also essential for TC climatology models, as TCs undergoing RI account for the majority of the secondary peak in lifetime maximum intensity distributions (Lee et al., 2016b). Although the PepC-Global intensity module reproduces this secondary peak, it is useful to verify whether the underlying mechanism is consistent with observations. TCs are divided into two groups based on whether they undergo RI, defined as at least a 30-knot intensity increase within a 24-hour window. As shown in Fig. 10, the PepC-Global intensity module reproduces distributions close to observations for both groups. However, it is worth noting that the discrepancy is most significant in ENP, where the simulated lifetime maximum intensity distributions of rapidly intensifying TCs exhibit a negative bias.

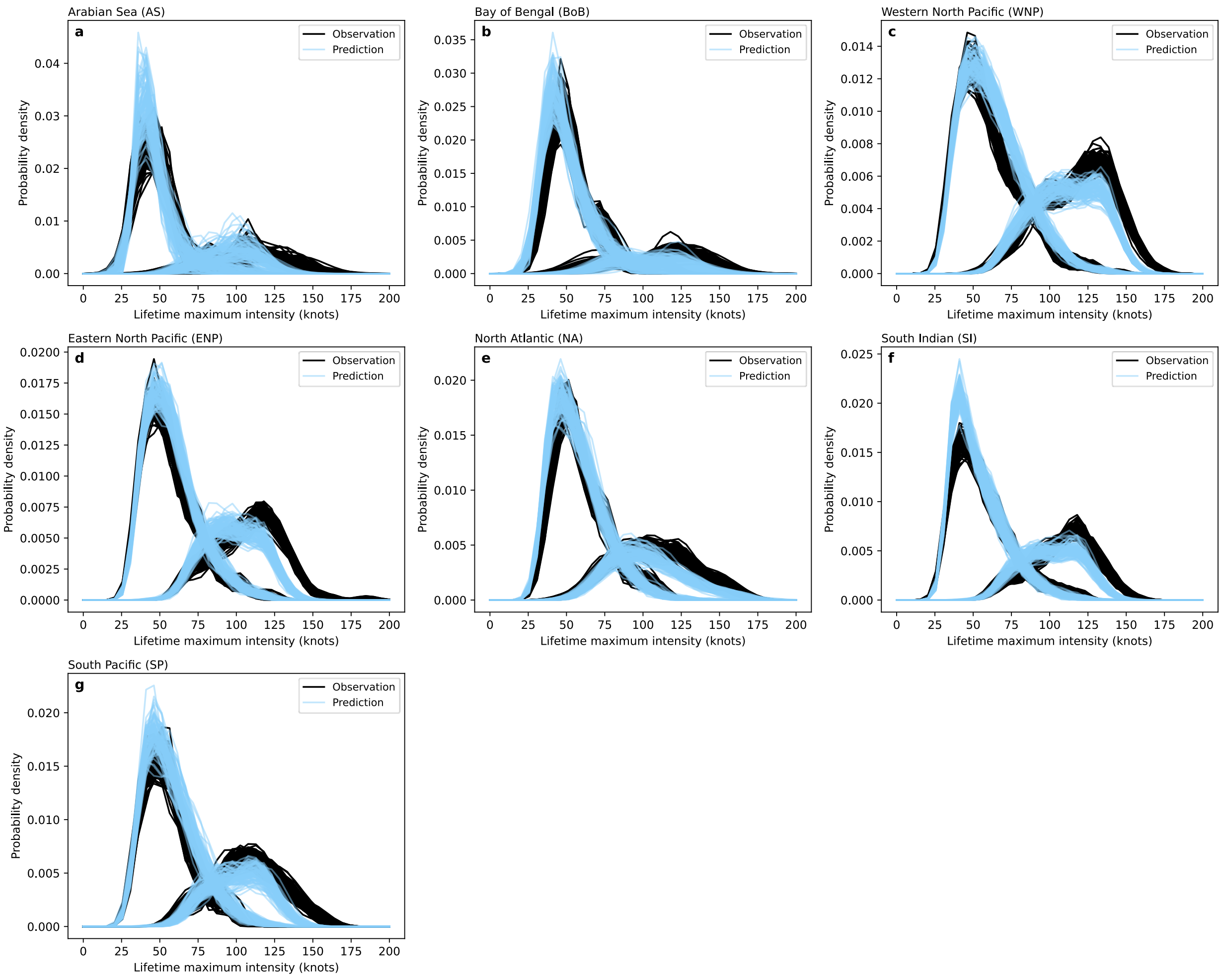

**Figure 10. Basin-wise evaluation of lifetime maximum intensity distributions for non–RI and RI TCs simulated by the PepC-Global intensity module along observed tracks.** Same as (a–g) in Fig. 9 but separated into two groups, including non-RI TCs (left) and RI TCs (right).

Landfall intensity is the most direct determinant of real coastal losses, making its accurate reproduction a critical test of TC climatology model skill. Accurately predicting landfall intensity requires not only reproducing the lifetime maximum intensity but also capturing the subsequent intensity decay as TCs approach the coast. When initialized and propagated along observed tracks, the PepC-Global intensity module may terminate a storm early (when intensity falls below 11 knots) before completing the observed full track or reaching landfall. In observations, 841 TCs make landfall along the five selected coastlines. Across the 100 realizations, 80,903 TCs make landfall, representing 96% of the expected total. For TCs that do not make landfall in the observations, tracks are not extended, and thus they also do not make landfall in the simulations. The 4% difference is partly attributable to this experimental design rather than solely to errors in the PepC-Global intensity module. Despite some discrepancies, the PepC-Global intensity module largely reproduces the observed landfall intensity distributions, as shown in Fig. 11. Compared with the performance for lifetime maximum intensity, the agreement for landfall intensity is somewhat lower, suggesting that landfall intensity is more difficult to predict than lifetime maximum intensity.

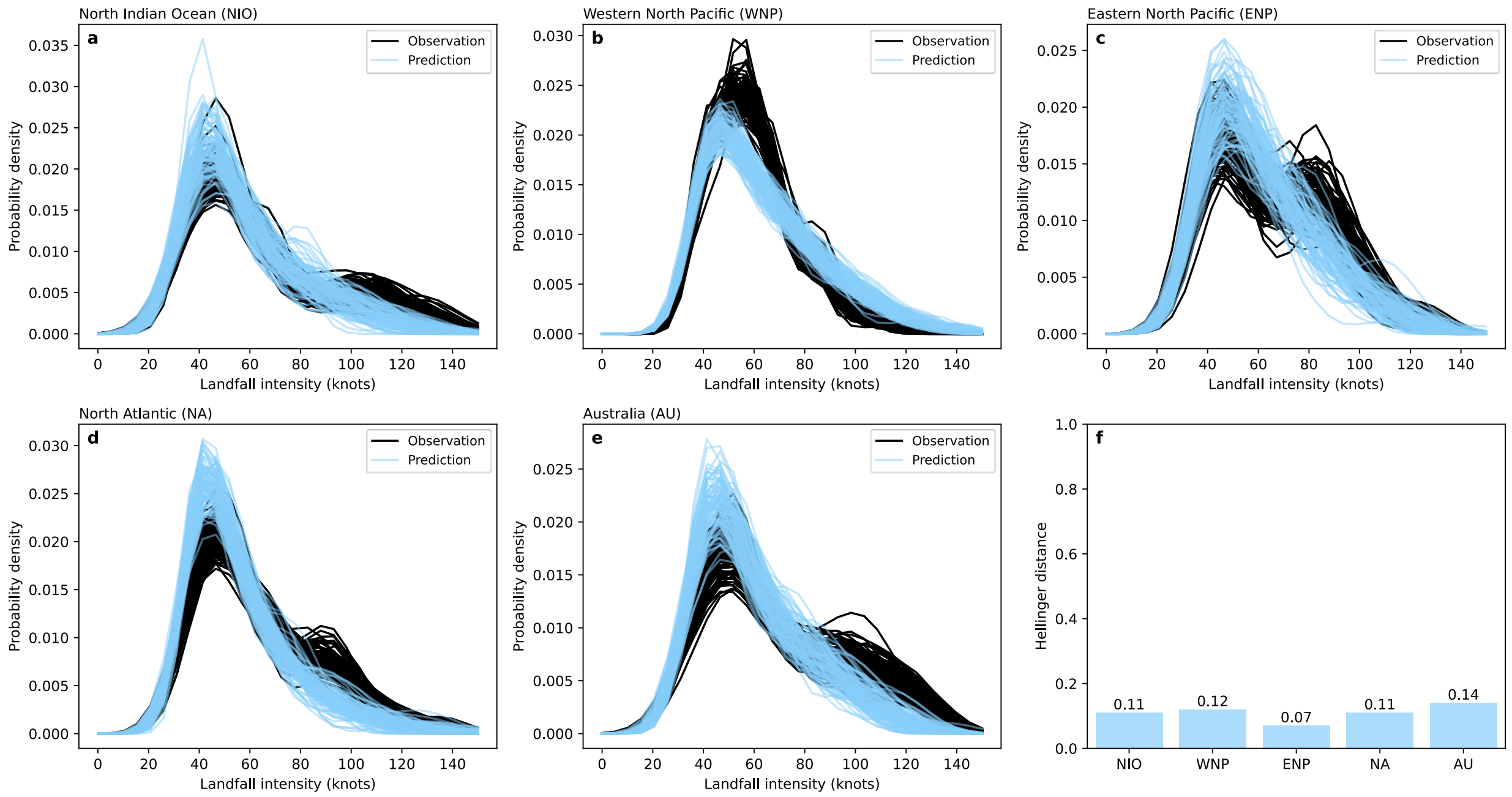

**Figure 11**. **Basin-wise evaluation of landfall intensity distributions simulated by the PepC-Global intensity module along observed tracks.** Same as Fig. 9 but for landfall intensity along the five selected coastlines. Only landfalls with intensity of at least 35 knots are included.

Landfall intensity varies considerably across locations. While accurate reproduction of basin-wide landfall intensity distributions is important, predicting landfall intensity at the local scale is even more critical for risk assessment. Because simulations are run along observed tracks, each observed landfall location can be paired with approximately 100 simulated intensities from the PepC-Global intensity module, enabling location-by-location probabilistic evaluation. As shown in Fig. 12, almost all observed landfall intensities fall within the 95% range of the corresponding simulated landfall intensities across basins. No landfall locations are identified as overestimated. Only along the Houston coastline in NA, the Andhra Pradesh coastline in NIO, and the Pilbara coastline near Port Hedland in Western Australia are landfall intensities identified as underestimated.

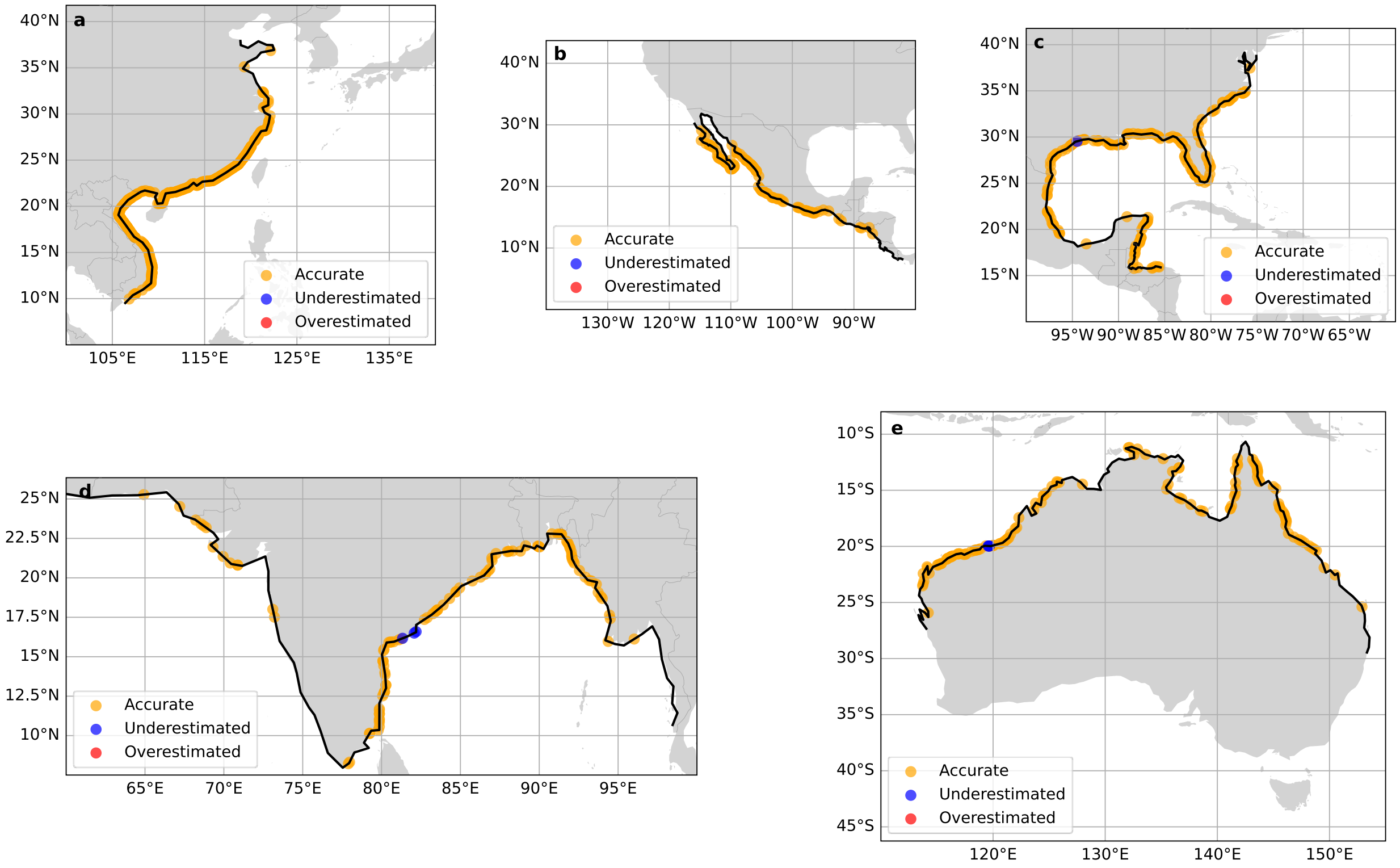

**Figure 12. Basin-wise evaluation of landfall intensity at observed landfall locations simulated by the PepC-Global intensity module along observed tracks.** In (a–e), dots mark observed landfall locations, with colors indicating whether the observed landfall intensity falls below the 2.5th percentile of the simulated distribution (overestimated; red), within the central 95% range (accurate; orange), or above the 97.5th percentile (underestimated; blue). At each observed landfall position, approximately 100 simulations of landfall intensity are produced from storms that remain above the termination threshold (11 knots) up to that point.

## 6. Integrated System Evaluation

Having evaluated the genesis, track, and intensity modules of PepC-Global separately, it is also important to evaluate the fully coupled system with all three components active. This tests how errors propagate from one module to another and reveals how imperfections in one module affect the performance of the others. When modules are evaluated separately, each receives observed inputs. In coupled mode, however, errors from upstream modules propagate to downstream modules, so the fully coupled system is expected to perform no better—and likely worse—than individual modules evaluated in isolation.

However, the fully coupled PepC-Global reproduces lifetime maximum intensity distributions more closely (Fig. 13 vs. Fig. 9). Compared to simulations along observed tracks (Fig. 9), simulations using simulated genesis and tracks show higher agreement with observations, particularly for WNP, ENP, and SI. The secondary peak and right tails align better with observations. Hellinger distances decrease in all basins, by 0.01–0.03 (Fig. 13h vs. Fig. 9h), confirming this improvement. The spread across 100 realizations shows no significant difference between simulations along observed tracks and those along simulated tracks. Why do simulations using simulated genesis and tracks perform better? Comparing Fig. 14 with Fig. 10 reveals that the ability to simulate rapidly intensifying TCs is improved, particularly for WNP, ENP, and SI. In summary, the fully coupled PepC-Global reproduces lifetime maximum intensity distributions more accurately, with comparable spread, and for the correct physical reason—improved representation of the secondary peak.

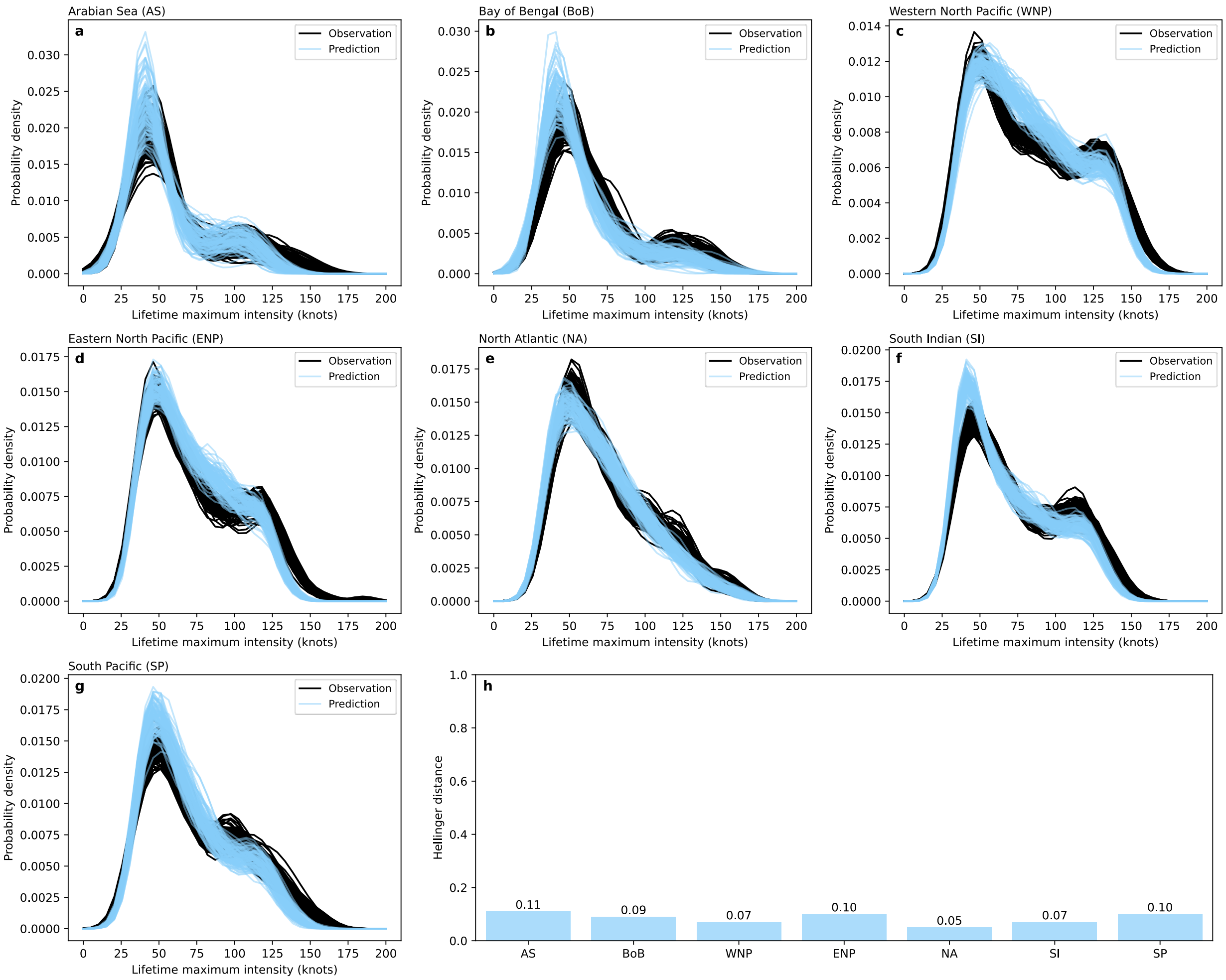

**Figure 13. Same as Fig. 9 but for the fully coupled PepC-Global.**

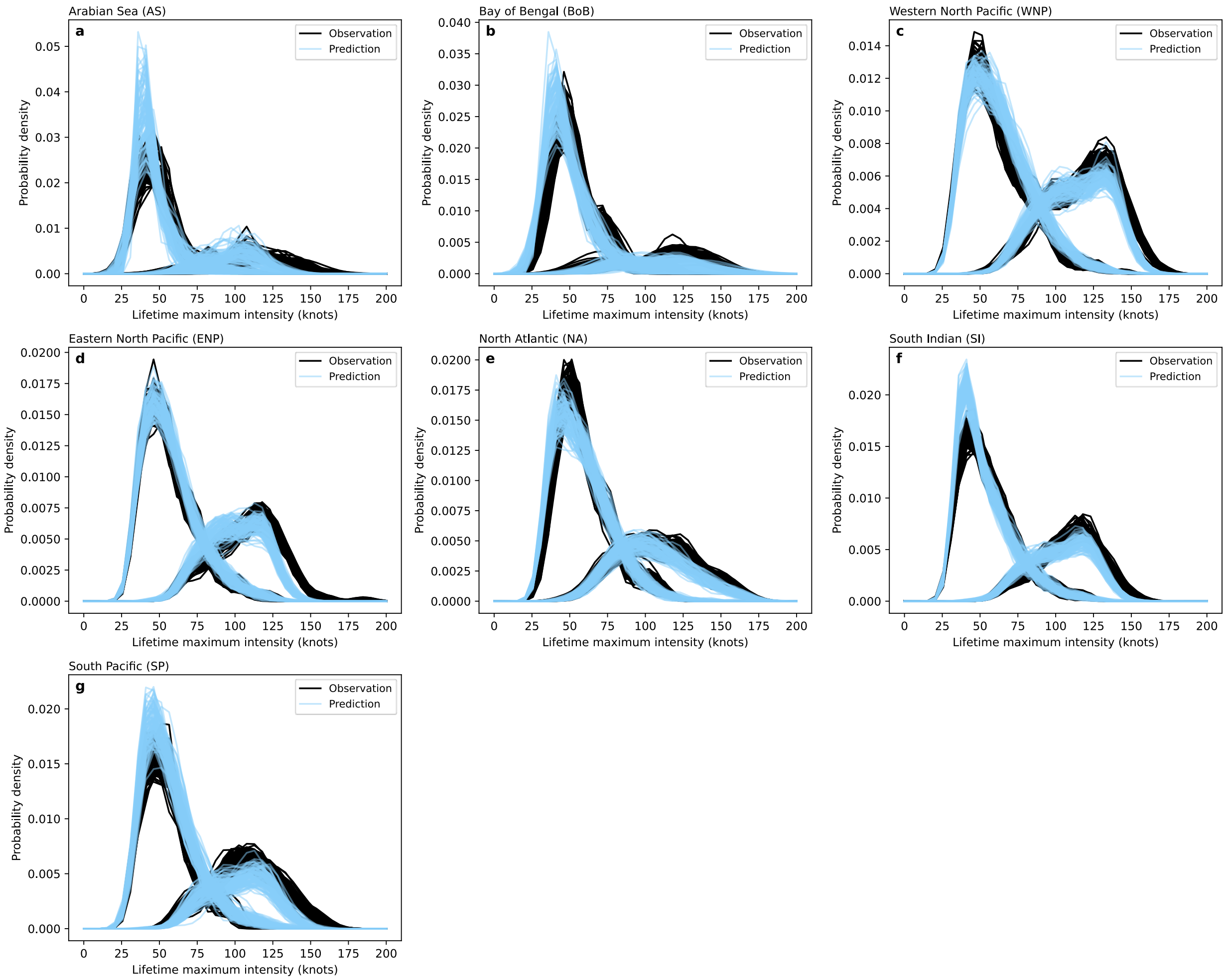

**Figure 14. Same as Fig. 10 but for the fully coupled PepC-Global intensity module simulations (I3).**

One plausible explanation is that observed TC tracks and intensities are the product of two-way interactions. When the intensity module is forced along observed tracks in a one-way configuration, this coupling is broken. The observed tracks are the final outcomes filtered by the actual intensity evolution, yet the intensity model cannot use this information directly. In the fully coupled system, the simulated tracks and intensities co-evolve, preserving this two-way interaction and allowing the intensity module to more naturally reproduce rapid intensification events. Meanwhile, the spread across realizations does not expand significantly because the stochastic variability introduced by the genesis and track modules is physically constrained rather than adding independent noise. The track-related uncertainty is already implicitly embedded in the intensity model stochastic terms, so coupling does not compound the sources of randomness but instead channels them through a more coherent physical framework.

Landfall tells a complex story. The fully coupled PepC-Global shows comparable skill in reproducing landfall frequency in WNP and NA, but reduced skill in other basins (Fig. 15 vs. Fig. 8). Conversely, the fully coupled PepC-Global shows increased skill in reproducing landfall intensity distributions in NA, slightly increased skill in NIO, slightly reduced skill in ENP, and comparable skill in WNP and AU (Fig. 16 vs. Fig. 11). Taken together, the fully coupled PepC-Global captures the key features of landfall climatology across basins, with only ENP showing slightly degraded performance in both frequency (relative to simulations from observed genesis) and intensity (relative to simulations along observed tracks).

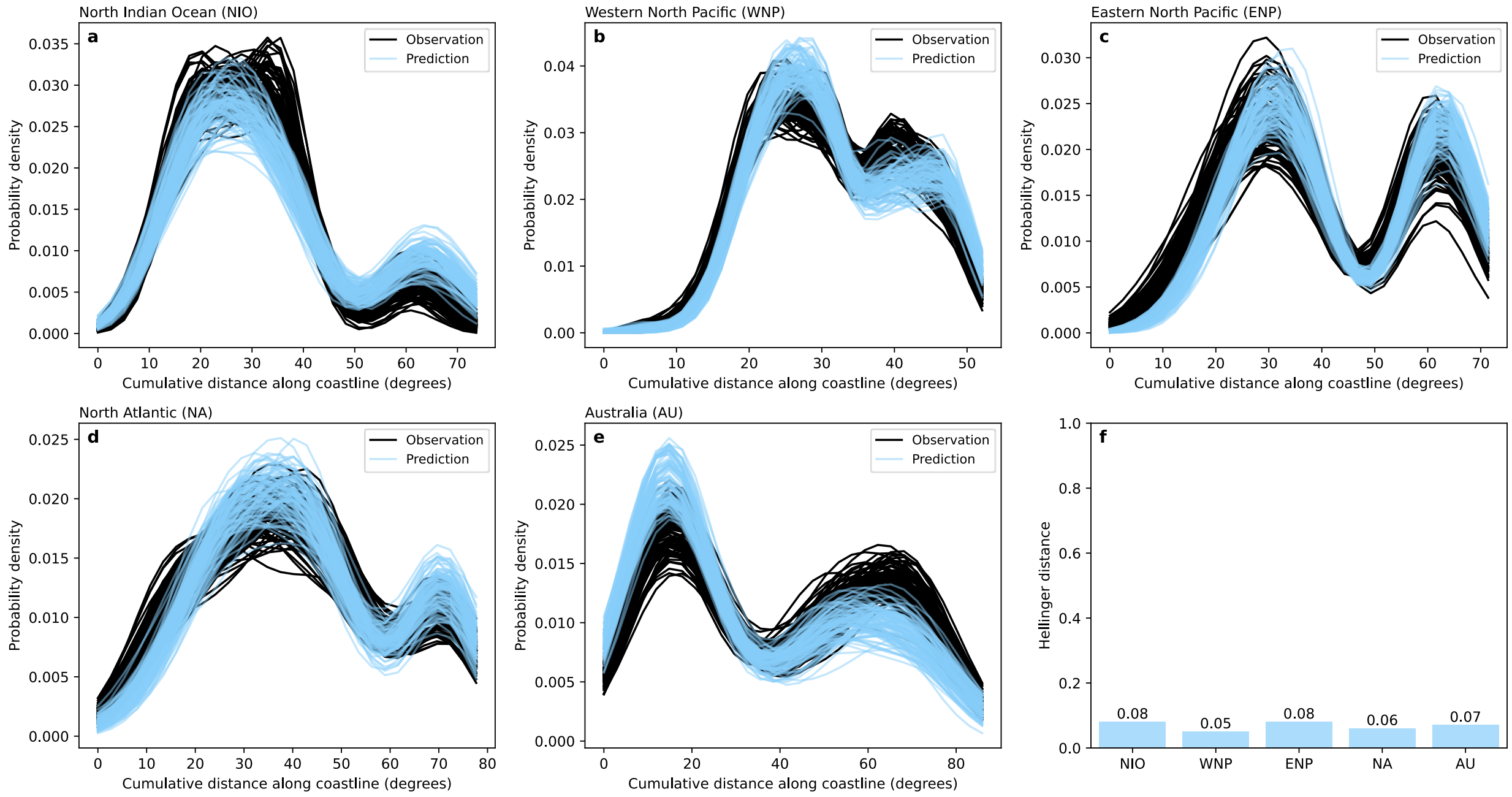

**Figure 15. Same as Fig. 8 but for the fully coupled PepC-Global.**

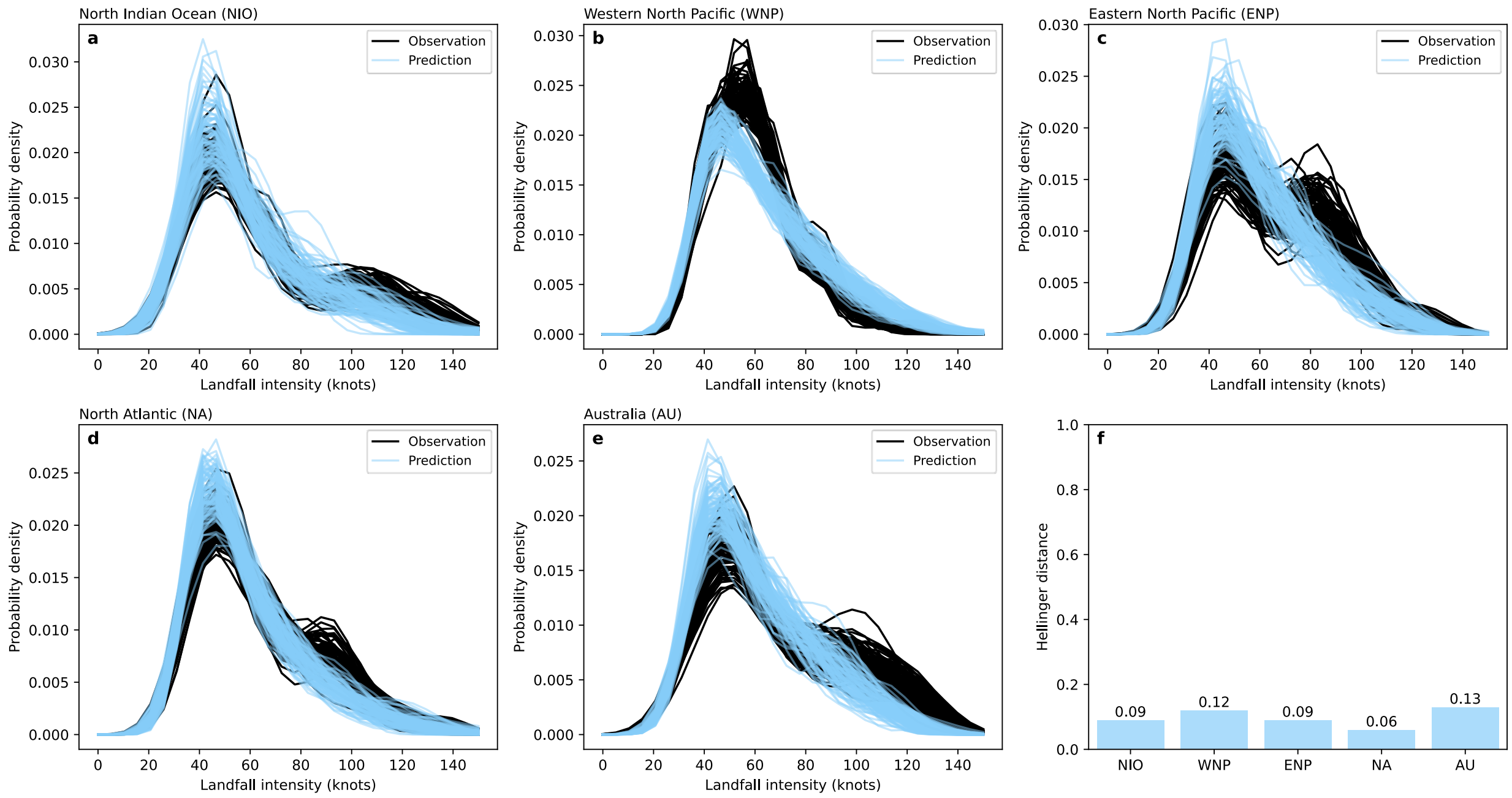

**Figure 16. Same as Fig. 11 but for the fully coupled PepC-Global.**

## 7. Summary and Discussion

We developed PepC-Global, a basin-tuned, environment-dependent probabilistic framework for global TC climatology that generates large synthetic catalogs by coupling stochastic genesis, track, and intensity modules. The genesis, track, and intensity modules outperform linear baseline models in out-of-sample evaluations, and the intensity module incorporates an environment-dependent over-land decay model, offering greater sensitivity to climate change signals than fixed decay rate approaches. Across major basins, PepC-Global reproduces observed features in genesis seasonality and interannual variability, spatial patterns of track passage, along-coastline landfall frequency, and basin-wise lifetime maximum intensity and landfall intensity probability densities. Systematic evaluations have demonstrated close agreement between PepC-Global simulations and observations. Overall, PepC-Global provides an end-to-end probabilistic basis for estimating TC hazard and risk and a practical framework for investigating changes in TC activity across future climate scenarios.

Future work will apply PepC-Global to climate-model output to quantify how TC hazards and associated risks may evolve under climate variability and anthropogenic forcing. Because PepC-Global is driven by large-scale environmental predictors at monthly and coarse spatial resolution, it is portable across a broad range of climate models and scenarios, enabling the generation of large synthetic TC ensembles and the estimation of changes in regional track passages, landfall frequency, and intensity exceedance probabilities that are difficult to sample directly from model-simulated storms. This approach also provides a pathway to attribute projected hazard changes to shifts in environmental controls on genesis, steering, and intensification, while separating uncertainty from internal stochasticity and intermodel spread. Such applications will facilitate consistent, basin-wide assessments of future TC hazard and risk for coastal impacts and catastrophe modeling.

Although PepC-Global is developed for climatological applications using monthly environmental predictors, it may be adaptable to other time scales. For seasonal forecasting, the framework could potentially be driven by seasonal climate predictions from dynamical models, enabling probabilistic TC activity forecasts several months in advance. For subseasonal prediction, incorporating higher-frequency environmental data might allow PepC-Global to capture intraseasonal variability such as the Madden–Julian Oscillation. However, such extensions would require careful validation to ensure that the statistical relationships learned from the historical period remain valid under different climate states or at different time scales. While these applications remain speculative, the modular design of PepC-Global provides a foundation for exploring such extensions in future work.

**Acknowledgments**

This work is supported by the National Science Foundation as part of the Megalopolitan Coastal Transformation Hub (MACH) under NSF award ICER-2103754. This is MACH contribution number 97. Cong Gao is additionally supported by a postdoctoral fellowship from the Gordon and Betty Moore Foundation.

## Data Availability Statement

The datasets used in this study are publicly available. ERA5 atmospheric variables and SST fields were obtained from the Copernicus Climate Data Store (CDS). Ocean subsurface fields were obtained from the NCEP GODAS hosted by NOAA Physical Sciences Laboratory (PSL). TC best-track data were obtained from the IBTrACS provided by NOAA NCEI.

ERA5 (CDS; hourly pressure levels): https://cds.climate.copernicus.eu/datasets/reanalysis-era5-pressure-levels?tab=overview

ERA5 (CDS; hourly single levels): https://cds.climate.copernicus.eu/datasets/reanalysis-era5-single-levels?tab=overview

ERA5 (CDS; monthly pressure-level means): https://cds.climate.copernicus.eu/datasets/reanalysis-era5-pressure-levels-monthly-means?tab=overview

ERA5 (CDS; monthly single-level means): https://cds.climate.copernicus.eu/datasets/reanalysis-era5-single-levels-monthly-means?tab=overview

GODAS (NOAA PSL): https://psl.noaa.gov/data/gridded/data.godas.html

IBTrACS (NOAA NCEI product page): https://www.ncei.noaa.gov/products/international-best-track-archive

110m coastline geometry (Natural Earth): https://www.naturalearthdata.com/downloads/110m-physical-vectors

## Conflict of Interest Disclosure

The authors declare no competing interests.